\documentclass[aps,pre,twocolumn,groupedaddress,amsmath,amssymb,preprint]{revtex4-2} 
\usepackage{graphicx}
\usepackage{bm}
\usepackage{soul}
\usepackage{color}
\usepackage[T1]{fontenc}
\begin{document}
\preprint{}


\title{Metamaterial two-sphere Newton's cradle}


\author{Simon A. Pope}
\email[]{s.a.pope@sheffield.ac.uk}
\affiliation{Department of Automatic Control and Systems Engineering, University of Sheffield, Mappin Street, Sheffield, S1 3JD, UK.}

\author{Oliver B. Wright}
 \affiliation{Hokkaido University, Sapporo, 060-0808, Japan}
 \affiliation{Graduate School of Engineering, Osaka University,
   2-1 Yamadaoka, Suita, Osaka, 565-0871, Japan}


\date{\today}

\begin{abstract}
Locally resonant metamaterials are among the most studied types of elastic/acoustic metamaterials, with significant research focused on wave propagation in a continuum of ``meta-atoms.'' Here we investigate the collision dynamics of two identical pendulum-suspended mass-in-mass resonators, essentially a two-sphere Newton's cradle, emphasizing the readily realizable scenario where the internal resonator frequency is much greater than the pendulum frequency. We first show that the dynamics of a collision can be described using effective parameters, similar to how previous metamaterials research has characterized wave propagation through effective material properties. Non-conventional collision dynamics\textemdash observed in two colliding mass-in-mass systems where one is initially at rest\textemdash include behaviors such as the moving sphere rebounding as if from a fixed wall while the other remains essentially stationary, the spheres coupling and moving forward in near-unison, and the spheres recoiling in opposite directions. These responses can be achieved by tuning the effective parameters. We demonstrate that these parameters can take on values that differ significantly from those in a conventional Newton’s cradle. Additionally, we investigate multiple collisions of the two spheres, revealing complex dynamics. This work paves the way for the development and study of new ``collision-based metamaterial'' structures.
\end{abstract}


\maketitle

\section{Introduction}

Metamaterials are engineered structures that exhibit effective properties not typically found in conventional materials \cite{liu_metamaterials_2011}. Some of the most widely discussed properties are those that affect wave propagation within the bulk of the metamaterial, such as density and various moduli in elastic materials and acoustics. Methods for designing such metamaterials, in which one or more of these effective properties is negative for a desired frequency band, are of particular interest \cite{banerjee_waves_2019} due to the role of negative effective values in fundamentally modifying behavior compared to conventional materials and enabling novel applications \cite{milton_cloaking_2006, guenneau_acoustic_2007}.

Wave propagation in elastic metamaterials is typically considered in the context of a continuum that is permanently elastically connected. A class of widely studied structures corresponds to mass-in-mass locally resonant metamaterials \cite{huang_negative_2009, lee2016origin, banerjee_waves_2019}, where a lattice of connected masses each has a local resonator attached. In most formulations of mass-in-mass systems, the resulting dynamics are described by a frequency-dependent effective mass. A sufficiently strong resonance creates a frequency band in which the effective mass is negative, resulting in a complex modulus that attenuates wave propagation within this band. Some research has focused on wave propagation in granular metamaterials, composed of closely packed discrete entities \cite{kim_review_2019}. The contact between these entities is inherently nonlinear, with individual entities resisting compression but separating under tension. Some studies have concentrated on pre-compressed systems that do not separate under tension forces smaller than the pre-compressing force. This includes a one-dimensional array of beads with local ``ring'' resonators \cite{gantzounis_granular_2013}, which showed a transmission bandgap typical of single-negative metamaterials. The effect of resonant inclusions on the transmission, reflection, and localization of energy has been investigated for a single bead, and subsequently extended to two and three beads in a sequence initially in contact \cite{kevrekidis_interaction_2013}.

Other related work on granular metamaterials has examined their response as the ratio of resonator mass to transmission medium mass tends to zero \cite{hadadifard_massmass_2021}. Separate work has also explored the generation and identification of stationary and traveling waves in such structures \cite{xu_traveling_2015, kevrekidis_traveling_2016, liu_breathers_2016, liu_strongly_2016, bonanomi_wave_2015}. A ``woodpile'' metamaterial design has also been investigated \cite{kim_impact_2017}, where the meta-atoms are not permanently elastically connected but stacked so that they can separate and collide under specific conditions. One common feature across this previous work is the focus on multiple collisions and their associated wave propagation. However, the fundamental processes involved in individual collisions in such metamaterials and their effect on the overall dynamics have not been extensively studied or compared to conventional collision dynamics.

The best-known demonstration of collisions between conventional entities is Newton's cradle. Although commonly seen as an executive toy, it still holds significance from a research perspective. Its relatively simple setup can be modified to test collisions between ``non-standard'' entities. An interesting example involves a modified Newton's cradle where the spheres are coated in fluid, creating a ``Stokes cradle'' \cite{donahue_stokes_2010a, donahue_stokes_2010b}. When using conventional Newton's cradle conditions (i.e., a single initially displaced sphere) for a three-sphere cradle, four different responses arise: the standard Newton's cradle (where the last sphere separates and the first two cluster), all three spheres cluster, the last two spheres cluster and move away from the initially displaced sphere, and all three spheres separate. These outcomes depend on the Stokes number and liquid coating thickness.

In this paper, we use a model of a frictionless two-sphere Newton's cradle, modified so that the colliding entities are hollow mass-in-mass resonators. We investigate the effect of the local resonators on the fundamental properties of individual collisions between the outer spherical shells in a readily realizable case where the internal resonator frequency is significantly greater than the pendulum frequency. We show that the calculated velocities at the moment of separation can be expressed in a form equivalent to conventional sphere collisions. From this, we derive effective coefficients of restitution and motion using an approach similar to those typically applied in continuum metamaterials. However, unlike previous approaches, we determine these effective parameters in the time domain, as they are strictly defined only at the moment the shells separate at the end of a collision. These effective parameters can take non-conventional values, including negative ones. We show that by varying the ratio of the resonator frequency to the fundamental ``compression frequency'' of the colliding outer masses and the energy stored in the local resonators, it is possible to achieve separation velocities that do not comply with the conventional conservation of momentum for the outer masses. We also elucidate the possible temporal evolutions of this two-sphere cradle.

\section{Theory and Modeling}

\subsection{Basic equations of motion}

\begin{figure}
  \includegraphics[width=1\columnwidth]{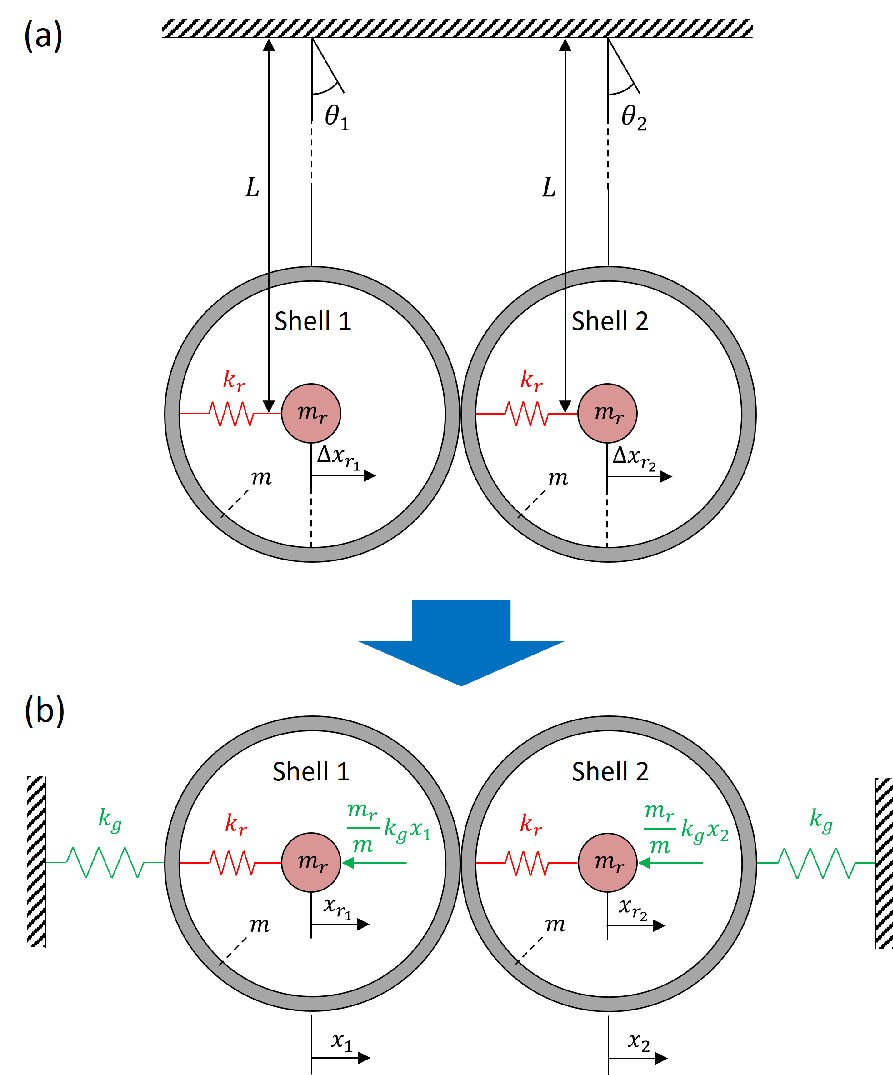}
 \caption{(a) The two-sphere mass-in-mass colliding pendulum system (a modified Newton's cradle), along with (b) the equivalent approximate pure mass-and-spring model applicable for small displacements and stiff spherical shells, which exhibits one-dimensional motion. The chosen model for the dynamics during a collision, which depends on the collision stiffness $k_c$ in a linear approximation, is not shown. In this paper, we assume that the right-hand shell and its internal mass are in their equilibrium positions before the first collision.}
\label{Figure1}
 \end{figure}

The colliding spheres are assumed to be two identical, suspended, frictionless mass-in-mass resonators, as shown in Fig.~\ref{Figure1}(a), which can swing in a single plane, resembling a two-sphere Newton's cradle. In a practical Newton's cradle configuration, a single suspension wire for each mass is replaced by two wires to ensure such motion. The spheres can be visualized as spherical shells, inside which a linear one-dimensional mass-spring resonator is attached. The angle $\theta_i$ ($i$=1, 2) is the angular displacement of the outer sphere $i$. The external shells are assumed to be suspended by identical wires of negligible mass attached to a fixed point above. This provides an external gravitational potential that acts to return both shells to their undisturbed positions, defined by the condition $\theta_{1,2}=0$. In these positions the external shells are positioned so that they are just touching, with their gravitational potential energy at its minimum. The solid internal masses are assumed to be spherical (although this is not a required assumption) and, for simplicity, are assumed to be in their central positions when in equilibrium. They are constrained to move perpendicular to the supporting wire, and $\Delta x_{r_i}$ is the component of the extension of a spring connected to its internal mass. The mass of the springs is assumed to be negligible compared to the other masses.

To simplify the analysis of this colliding mass-in-mass system, a one-dimensional motion approximation is applied. This is illustrated in Fig.~\ref{Figure1}(b), where $x$ ($x_r$) represent the relative displacements of the outer spherical shells (internal resonators) from their equilibrium positions. As demonstrated in the Supplemental Material, under the assumption of small displacements, the effect of gravity on the angular motion of the spherical shells can be approximated by modeling the wires as linear springs, with a spring constant given by $k_g = mg/L$. Here, $m$ represents the mass of the spherical shells, $g$ is the acceleration due to gravity, and $L$ is the effective length of the pendulums, measured from the support to the center of the shells. This represents the gravitational potential relative to the undisturbed position of each shell (which is taken as zero for an undisturbed shell). In Fig.~\ref{Figure1}(b), gravitational effects are indicated in green, while the effect of gravity on an internal mass is represented as an external force, given by $(m_r/m)k_g x_{r_i}$, since it cannot be depicted solely by a spring attached to the internal mass. The resulting resonance frequency is given by $\omega_g=\sqrt{k_g/m}=\sqrt{g/L}$. The mass and spring constant of the internal linear resonators are defined as $m_r$ and $k_r$, respectively, with the associated angular resonance frequency $\omega_r=\sqrt{k_r/m_r}$. The motion is assumed to be lossless. The present work does not focus on exploring various resonator designs that could physically realize the desired dynamics. Instead, we use the simple representative example shown in Fig.~\ref{Figure1}(b), featuring spherical masses, to investigate the fundamental dynamics of such systems.

The equations of motion for the two shells with internal resonators, which oscillate freely in a one-dimensional potential field until they come into contact with one another, can be modeled as a piecewise system. A similar approach has previously been used to model the conventional Newton's cradle \cite{hutzler_rocking_2004}. To ensure that the contact force only acts under compression (i.e., that no forces act to resist tension), we use Eqs.~(\ref{eq:contact}) to describe the motion when the shells are in contact for $x_2-x_1\leq0$ and Eqs.~(\ref{eq:pendulum}) (describing pendulum dynamics) when the shells are not in contact:
\begin{subequations}
\begin{eqnarray}
&\text{when } x_2-x_1\leq0\text{,} \nonumber \\
&m\ddot{x}_1=k_c\left(x_2-x_1\right)-k_g x_1+k_r\left(x_{r_1}-x_1\right) \label{eq:contact_1a},\\
&m_r\ddot{x}_{r_1}=-\frac{m_{r}}{m} k_g x_1 + k_r\left(x_1-x_{r_1}\right) \label{eq:contact_1b}, \\
&m\ddot{x}_2=-k_c\left(x_2-x_1\right)-k_g x_2+k_r\left(x_{r_2}-x_2\right) \label{eq:contact_2a},\\
&m_r\ddot{x}_{r_2}=-\frac{m_{r}}{m} k_g x_2 + k_r\left(x_2-x_{r_2}\right) \label{eq:contact_2b},
\end{eqnarray}
\label{eq:contact}
\end{subequations}
\begin{subequations}
\begin{eqnarray}
&\rm{otherwise,} \nonumber \\
&m\ddot{x}_1=-k_g x_1+k_r\left(x_{r_1}-x_1\right), \\
&m_r\ddot{x}_{r_1}=-\frac{m_{r}}{m} k_g x_1 + k_r\left(x_1-x_{r_1}\right), \\
&m\ddot{x}_2=-k_g x_2+k_r\left(x_{r_2}-x_2\right), \\
&m_r\ddot{x}_{r_2}=-\frac{m_{r}}{m} k_g x_2 + k_r\left(x_2-x_{r_2}\right).
\end{eqnarray}
\label{eq:pendulum}
\end{subequations}\\
During contact, the force between the spherical shells is assumed to be compressive. Contact forces result from the relatively small indentation of the shells in the region around the contact point, and for purposes of approximation we ignore in Eqs.~(\ref{eq:contact}) the Hertzian-contact nonlinearity and assume this compressive force to be linear in indentation, as was done previously for the case of contacting spherical shells \cite{hertz_contact_1881, landau_theory_1986, herrmann_how_1982}. The contact stiffness of the shell is defined as $k_c=O\left(Eh^2/2R\right)$, where $E$ is the Young's modulus of the shell material, $h$ is the thickness of the shell, and $R$ is the shell outer radius. In contrast, the indentation of solid spheres requires precise treatment using (non-linear) Hertzian theory \cite{johnson1987contact}, combined with the theory of shell deformation. We also define a characteristic compression angular frequency $\omega_c=\sqrt{2k_c/m}$, which is useful in the elucidation of the contact dynamics.

Equations~(\ref{eq:contact}) represent a set of unforced equations governing the contact dynamics. During a collision, the initial conditions of Eqs.~(\ref{eq:contact}) are defined by the state of the pendulum dynamics given by Eqs.~(\ref{eq:pendulum}) at the time of contact $t_c$ defined when $\left(x_2-x_1\right)_{t_c}=0$. For simplicity of the analysis throughout this paper, the initial conditions of the shells at the start of a collision are taken as those of a  conventional Newton's cradle; the first shell moves and collides with the second shell, which, along with its internal resonator, is initially stationary at the equilibrium position (i.e., having zero potential and kinetic energy). This gives $x_{1,t_c}=x_{2,t_c}=\dot{x}_{2,t_c}=0$ and $\dot{x}_{1,t_c}\neq0$. The contact period is defined as the period $[t_c, t_s]$ between the shells making contact at time $t_c$ and then separating at time $t_s$, that is, the period during which $\left(x_2-x_1\right)\leq0$. The state of the system at the end of the contact period (i.e., at $t_s$) then defines the initial conditions for the pendulum dynamics governing the motion after separation. 

\subsection{Derivation of the effective parameters and determination of the post-collision velocities}

The coefficient of restitution ($\mathit{CR}$) is one of the most commonly used quantities used to describe the collision between two bodies. It is usually defined in terms of the ratio of the relative separation to relative collision velocity:
\begin{equation}
CR = \left(\dot{x}_2-\dot{x}_1\right)_{t_s}/\left( \dot{x}_1-\dot{x}_2\right)_{t_c}.
\label{eq:CR}
\end{equation}
The post-collision velocities for a simple collision between two hard, solid spheres, such as in a basic model of a conventional Newton's cradle, where the second sphere is initially at rest ($\dot{x}_{2,t_c} = 0$) and where momentum is conserved and assumed to be transferred instantaneously, can be written in the following form:
\begin{subequations}
\begin{eqnarray}
&\dot{x}_{1,t_s} = \frac{m_1 - m_{2}CR}{m_1+m_2}\dot{x}_{1,t_c}, \\
&\dot{x}_{2,t_s} = \frac{m_1 + m_{1}CR}{m_1+m_2}\dot{x}_{1,t_c}.
\end{eqnarray}
\label{eq:v1_and_v2_CR_conv}
\end{subequations}
For identical spheres this reduces to
\begin{subequations}
\begin{eqnarray}
&\dot{x}_{1,t_s} = \frac{1}{2} \left(1 - CR\right)\dot{x}_{1,t_c} \label{eq:v1_CR_conv_identical}, \\
&\dot{x}_{2,t_s} = \frac{1}{2} \left(1 + CR\right)\dot{x}_{1,t_c} \label{eq:v2_CR_conv_identical}.
\end{eqnarray}
\label{eq:v1_and_v2_CR_conv_identical}
\end{subequations}\\
The coefficient of restitution $CR$ represents how well energy is conserved in the collision. $CR$$=$1 for a conventional two-sphere Newton's cradle with linear lossless compression during contact, which represents a perfectly elastic collision. The ``1'' in Eqs.~(\ref{eq:v1_and_v2_CR_conv_identical}) arises owing to the conservation of momentum: addition of Eqs.~(\ref{eq:v1_CR_conv_identical}) and (\ref{eq:v2_CR_conv_identical}) to eliminate $\mathit{CR}$ leads to the momentum conservation equation $\dot{x}_{1,t_c} = \dot{x}_{1,t_s} + \dot{x}_{2,t_s}$ for identical spheres. Using this understanding, we can write a more general form of Eqs.~(\ref{eq:v1_and_v2_CR_conv_identical}) as follows:
\begin{subequations}
\begin{eqnarray}
&\dot{x}_{1,t_s} = \frac{1}{2}\left(\mathit{CM}_e-\mathit{CR}_e\right)\dot{x}_{1,t_c}, \\
&\dot{x}_{2,t_s} = \frac{1}{2}\left(\mathit{CM}_e+\mathit{CR}_e\right)\dot{x}_{1,t_c}.
\end{eqnarray}
\label{eq:v1_and_v2_CR_effective}
\end{subequations}\\
 Here we define an effective coefficient of restitution in the same manner as the conventional sense:
\begin{subequations}
\begin{eqnarray}
&\mathit{CR}_e=\frac{\dot{x}_{2, t_s}-\dot{x}_{1, t_s}}{\dot{x}_{1, t_c}} \label{eq:CR_e_def}, \\
&\mathit{CM}_{e}=\frac{\dot{x}_{1, t_s}+\dot{x}_{2, t_s}}{\dot{x}_{1, t_c}} \label{eq:CM_e_def}.
\end{eqnarray}
\end{subequations}\\
We also define an effective ``coefficient of motion'' term in Eqs.~(\ref{eq:CM_e_def}) as the ratio of the momentum after to that before the collision of two identical spheres. When momentum is conserved, $\mathit{CM}_e = 1$. We can now use this more general definition to represent the collision of mass-in-mass spheres, and understand how the internal resonators influence the resulting post-collision motion. We refer to these variables as ``effective'' coefficients, since we will show that they serve to characterize the dynamics of the two colliding locally-resonant spheres, and are a function of the internal resonator parameters. This ``effective systems approach'' is analogous to that previously used to describe wave propagation in mass-in-mass continuum metamaterials.

To determine the effective coefficients of motion and restitution for the collision of such mass-in-mass entities, one needs to consider the velocities of both the spherical shells and the inner masses at the moment of separation. By taking into account the initial conditions at the moment of collision and making use of Eqs.~(\ref{eq:contact}) in the Laplace domain, one can eliminate $x_{r_1}$ and $x_{r_2}$ to obtain equations for $\dot{x}_{2, t_s}-\dot{x}_{1, t_s}$ and $\dot{x}_{2, t_s}+\dot{x}_{1, t_s}$ for purposes of substitution into Eqs.~(\ref{eq:CR_e_def}) and (\ref{eq:CM_e_def}), respectively, under the additional assumption of stiff shells, i.e., $\omega_g<<\omega_c$ (see the Supplemental Material and estimates in the main text below):
\begin{widetext}
 \begin{subequations}
\begin{eqnarray}  
&\mathit{CR}_e =-\left.\mathcal{L}^{-1} \left(\frac{\hat{m}_{e,s}s}{\hat{m}_e s^2+\omega_c^{2}}\left(\frac{x_{r_1,t_c}}{\dot{x}_{1, t_c}}s+\frac{\dot{x}_{r_1,t_c}}{\dot{x}_{1, t_c}}\right) +\frac{s}{\hat{m}_e s^{2}+\omega_c^{2}}\right)\right|_{t=t_s},
\label{eq:CR_effective}   \\
&\mathit{CM}_e =\left.\mathcal{L}^{-1}\left(\frac{\hat{m}_{e, s}}{\hat{m}_{e} s}\left(s \frac{x_{r_1,t_c}}{ \dot{x}_{1, t_c}}+\frac{\dot{x}_{r_1,t_c}}{ \dot{x}_{1, t_c}}\right)+\frac{1}{\hat{m}_{e} s}\right)\right|_{t=t_{s}},  
\label{eq:CM_effective}   \nonumber \\ 
\end{eqnarray}   
\end{subequations}
\text{where} \nonumber \\
 \begin{subequations}
\begin{eqnarray}
&\hat{m}_{e}=1+\frac{m_r}{m} \frac{\omega_r^{2}}{s^{2}+\omega_r^{2}}=1+\hat{m}_{e,s}.  \nonumber
\end{eqnarray}
\end{subequations}
These equations describe the effective coefficients of restitution and motion, respectively, where $\mathcal{L}^{-1}$ denotes the inverse Laplace transform. For ease of analysis, the time domain has been shifted such that $t_c = 0$. In this paper, we focus on the case in which the local resonant frequency is commensurate with the compression frequency, i.e., $\omega_r \sim \omega_c$, as this condition represents the one most practically accessible and where the effect of the internal resonators becomes most pronounced.

\end{widetext}
The result is that the effective coefficient of restitution is a function of two angular frequencies resulting from 1) the interaction during contact of the spherical shells, characterized by $\omega_c = \sqrt{2k_c/m}$, and 2) the local-resonance dynamics, charcterized by $\omega_r = \sqrt{k_r/m_r}$, whereas the effective coefficient of motion is a function of $\omega_r$ only. Both effective coefficients are a function of the resonator conditions at the moment of collision.

It is pertinent at this point to justify the above assumptions for a practical situation by giving order of magnitude estimates of a possible configuration for experimental realization of a two-sphere mass-in-mass Newton's cradle. This also helps to visualize an actual future experimental  configuration. Reasonable values for a possible implementation are as follows: steel shells of outer radius $R$=10~mm and wall thickness 1~mm, and inner tungsten spheres of radius 5~mm, leading to masses $m$=8.9~g, $m_r$=10.1~g, gravitational spring constant $k_g=0.87$~N/m, contact stiffness $k\approx 9.5\times10^6$~N/m, making use of the Young's modulus of steel $E=190$~GPa, density of steel 7850~kg~m$^{-3}$, density of tungsten 19250~kg~m$^{-3}$ and wire length $L$=100~mm; one can, for example, choose the inner-spring constant value of $k_r=2\times10^6$~N/m, in line with commercially available stiff coil springs of length ~5~mm. The resulting resonant frequencies correspond to compression frequency $\omega_c/2\pi=$7340~Hz, local-resonator frequency $\omega_r/2\pi=2240$~Hz and pendulum-resonance frequency $\omega_g/2\pi=1.6$~Hz. Clearly, this justifies the stiff shell approximation, $\omega_g \ll \omega_c$, and the assumption that $\omega_r$$\sim$$\omega_c$, both of which will be applied in the subsequent analysis. The equivalent conditions for the spring stiffness are  $\sqrt{k} \sim \sqrt{mk_r/2m_r}$. and $\sqrt{k_g}\ll\sqrt{2k}$. This ensures that the local resonator frequency is significantly lower than the resonant frequency of acoustic waves within the sphere, maintaining the validity of the lumped element approximation. Furthermore, we neglect the effect of acoustic waves generated in the wire by the high-frequency internal oscillator.

By increasing the shell radius $R$ while maintaining its thickness at $0.1R$, the compression frequency is reduced\textemdash enabling a more practical physical implementation using resonator springs with lower stiffness and greater length. The compression frequency can also be decreased by selecting a softer shell material (e.g., acrylic). We assume that the characteristic angular frequencies $\omega_g$, $\omega_c$ and $\omega_r$ are all well removed from any natural resonant frequencies of the individual shells (such as than of the $n=2$ flexural mode).

We carried out simulations with softer shells (resulting in longer contact times) and material losses (non-elastic collisions) to confirm that the subsequent analysis remains valid unless the choice of elastic parameters results in the above inequalities involving the characteristic frequencies being violated. The internal stiffness could be practically implemented with, for example, a commercially available coil spring. The internal mass's motion could be constrained to the collision's normal direction using a drilled mass sliding along a guiding rod attached to the shell's interior. Alternatively, leaf springs could provide a sufficient stiffness together with directional constraint.

The initial conditions of the internal mass in shell 1 (e.g., its potential and kinetic energy) could be set manually by pulling a rod extending from a hole in the shell opposite to the collision surface\textemdash and/or through more complex electromechanical methods, such as electronically releasing a pre-loaded resonator spring just before a collision. 

Importantly, the derivations and analysis presented are valid for collisions occurring normal (i.e., perpendicular) to the surface of any body, provided the underlying assumptions hold. For instance, the collision of two trolleys containing internal resonators on a track curved in the vertical direction could serve as an alternative experimental validation of the principles discussed.

The effective coefficients as described above are written as a function of the same effective mass that has been the focus of previous work on mass-in-mass continuum metamaterials \cite{huang_negative_2009}. However, the effective coefficients that we are focusing on here are defined in the time domain, as opposed to the frequency domain usually considered in metamaterials research. The reason for using the time domain in this case is that the frequency domain is less intuitive, as the effective parameters are defined at a single moment in time, i.e., when the shells separate. These parameters are considered effective because they encapsulate both the effects of collision dynamics and the influence of the resonators' motion on the overall movement of the spherical shells. Under our approximations, the duration of the collisions is governed by unforced dynamics, determined solely by the individual mass-in-mass system parameters and the associated initial conditions. The final state of the impact occurs when the two entities separate, at which point the governing dynamics change. Use of the time domain rather than the frequency domain makes the analysis slightly less straightforward owing to the dependence of Eqs.~(\ref{eq:CR_effective}) and (\ref{eq:CM_effective}) on the moment of separation $t_s$. The moment of separation also depends on the interaction during contact of the resonances at $\omega_c$ and $\omega_r$  and on the resonator initial conditions, and is defined as the moment corresponding to the first instance when ${x}_{2}-{x}_{1}=0$ after time $t_c$. As a consequence of this complexity, a simple analytical solution to the effective coefficients does not exist, and numerical solutions are required.

\subsection{Derivation of effective parameters for general motion in the external potential post-collision}

The instantaneous velocities of the two spherical shells, $\dot{x}_{1,t_s}$ and $\dot{x}_{2,t_s}$, at the moment of separation do not completely define the motion after separation. This can be seen by solving Eqs.~(\ref{eq:pendulum}) for $\dot{x}_{1}$ and $\dot{x_2}$ at $t_s$ and using initial conditions for the shell and internal resonator motion at the moment of separation. Since the compression angular frequency $\omega_c$ is, in practice, that appropriate to stiff colliding shells, $\omega_g<<\omega_c$, and, owing to the relatively weak pendulum spring stiffnesses in a Newton's cradle setup, $\omega_g<<\omega_r$. The velocity of a sphere after separation can be derived as follows (see the Supplementary Material for the full derivation):
\begin{eqnarray}
\dot{x}_{i}=&-\frac{m x_{i,t_s} + m_r x_{r_i,t_s}}{m+m_r} \omega_g\sin{\left(\omega_g t_{s^+}\right)} \nonumber \\
&+ \frac{m \dot{x}_{i,t_s} + m_r \dot{x}_{r_i,t_s}}{m+m_r}\cos{\left( \omega_g t_{s^+}\right)} \nonumber \\
&-m_r\frac{x_{i,t_s} - x_{r_i,t_s}}{m+m_r}\sqrt{\frac{m+m_r}{m}}\omega_r\sin{\left(\sqrt{\frac{m+m_r}{m}}\omega_rt_{s^+}\right)} \nonumber \\
&+ m_r \frac{\dot{x}_{i,t_s} - \dot{x}_{r_i,t_s}}{m+m_r}\cos{\left(\sqrt{\frac{m+m_r}{m}}\omega_rt_{s^+}\right)},
\label{eq:pendulum_after_collision}
\end{eqnarray}
where $t_{s^+}$ is the time after separation, namely $t_{s^+}=t-t_s$. 
The post-collision velocity is therefore a sum of oscillatory terms at two angular frequencies. One of these is the pendulum angular frequency $\omega_g$ and the other the much higher mass-weighted local-resonance angular frequency $\omega_r\sqrt{(m+m_r)/m}$. Assuming that the magnitude of the higher frequency oscillation is sufficiently small so that the shells do not collide again shortly after they separate, the general motion in the external potential is determined by terms involving $\omega_g$ in Eqs.~(\ref{eq:pendulum_after_collision}). This assumption is usually satisfied when $\omega_g<<\omega_r$, because the magnitude of the displacement of the mass-weighted local resonance term is proportional to $1/\omega_r$. Under such conditions, the total instantaneous post-collision velocities defined in Eqs.~(\ref{eq:pendulum_after_collision}) can be simplfied to the following form by setting $t_{s^+}=0$: 
 \begin{eqnarray}
\dot{x}_{i,t_s}&=&\frac{m \dot{x}_{i,t_s} + m_r \dot{x}_{r_i,t_s}}{m+m_r} + m_r \frac{\dot{x}_{i,t_s} - \dot{x}_{r_i,t_s}}{m+m_r} \nonumber \\
&=&\dot{x}_{i,a}+\dot{x}_{i,r}.
\label{eq:pendulum_after_collision_inst}
\end{eqnarray}
The first term is the contribution from the motion in the external potential and the second is a contribution from the resonator causing the sphere to oscillate with respect to the motion in the general external potential.
 
We can therefore define two sets of effective collision parameters. One set describes the total instantaneous velocity of the sphere at the moment of separation, as outlined by Eqs.~(\ref{eq:CR_e_def})-(\ref{eq:CM_e_def}). The other set characterizes the instantaneous effect of the collision on the general motion in the external potential after the collision, defined by the initial velocities, $\dot{x}_{i,a}$, of the oscillatory motion associated with the resonant frequency of the external potential. The effective coefficient of restitution for the general motion in the external potential can be defined in the same manner as the conventional coefficient of restitution, i.e., $\it{CR}_{a} = \left(\dot{x}_{2,a}-\dot{x}_{1,a}\right)/\left( \dot{x}_{1,t_c}-\dot{x}_{2,t_c}\right)$. Taking the same initial conditions at the point of contact, namely $\dot{x}_{2,t_c}=0$, and substituting for $\dot{x}_{1,a}$ and $\dot{x}_{2,a}$ from Eqs.~(\ref{eq:pendulum_after_collision_inst}), we obtain

\begin{subequations}
\begin{eqnarray}
\it{CR}_{a} =& \frac{1}{m+m_r}&\frac{m\left(\dot{x}_{2,t_s}-\dot{x}_{1,t_s}\right)+m_r\left(\dot{x}_{r_2,t_s}-\dot{x}_{r_1,t_s}\right)}{\dot{x}_{1,t_c}}, \nonumber \\
=& \frac{1}{m+m_r}&\left(m\it{CR}_e + m_r\it{CR}_{r}\right) \label{eq:CR_effective_a}, \\ \nonumber \\
\it{CM}_a =& \frac{1}{m+m_r}&\frac{m\left(\dot{x}_{2,t_s}+\dot{x}_{1,t_s}\right)+m_r\left(\dot{x}_{r_2,t_s}+\dot{x}_{r_1,t_s}\right)}{\dot{x}_{1,t_c}} ,\nonumber \\
=& \frac{1}{m+m_r}&\left(m\it{CM}_e + m_r\it{CM}_{r}\right).\label{eq:m_effective_a}
\end{eqnarray}
\end{subequations}
The effective coefficient of restitution for the general motion in the external potential is thus given by a mass-weighted sum of two effective coefficients of restitution: the instantaneous effective coefficient of restitution and a new effective coefficient of restitution arising from the resonators. An effective coefficient of motion $\mathit{CM}_a$ for the general motion can be derived in a comparable manner, leading to Eqs.~(\ref{eq:m_effective_a}). 
Similarly, for the velocities,
\begin{subequations}
\begin{eqnarray}
\dot{x}_{1,a} =& \frac{1}{2}\left(\it{CM}_a-\it{CR}_a\right)\dot{x}_{1,t_c}, \\
\dot{x}_{2,a} =& \frac{1}{2}\left(\it{CM}_a+\it{CR}_a\right)\dot{x}_{1,t_c}.
\end{eqnarray}
\label{eq:x_effective_a}
\end{subequations}\\
To determine the additional effective collision parameters $\it{CR}_{r}$ and $\it{CM}_{r}$, the velocity of the resonator masses at the moment of separation needs to be included. The Supplemental Material contains the derivation of these terms and the effective coefficients for general velocities. The resulting effective coefficient of restitution and motion in this case can be expressed in the form 
\begin{widetext}
\begin{subequations}
\begin{eqnarray}
&\textit{CR}_{a} = \frac{m}{m+m_r}\left(CR_{e} + \left.\left(\hat{m}_{e,s}(t) * CR_{e}(t)\right)\right|_{t=t_{s}}-\left.\omega_r^2\left(\frac{d^{2} \hat{m}_{e,s}(t)}{d t^{2}}\frac{x_{r_1, t_{c}}}{\dot{x}_{1, t_c}}+\frac{d \hat{m}_{e,s}(t)}{d t}\frac{\dot{x}_{r_1, t_{c}}}{\dot{x}_{1, t_c}}\right)\right|_{t=t_{s}}\right),
\label{eq:CR_a_effective} \\
&\textit{CM}_{a} = \frac{m}{m+m_r}\left(1+\frac{m_r}{m}\frac{\dot{x}_{r,t_c}}{ \dot{x}_{1, t_c}}\right).
\label{eq:CM_a_effective}
\end{eqnarray}
\end{subequations}
\end{widetext}
These equations are written in the time domain, where `$(t)$' denotes the time domain form of the equations for the effective mass and coefficient of restitution of the instantaneous velocity prior to their calculation at the moment of separation.

\section{Results and Discussion}

\subsection{Parametric studies of the dynamics of a particular system}

To aid analysis and without significant loss of generality, we consider the particular case where the resonator mass is equal to the spherical-shell mass, i.e., $m_r = m$. We define $\Omega = \omega_r / \omega_c$, so for a constant $\omega_c$, determined by the shell material, one can consider variations in $\Omega$ for the case of a constant resonator mass $m_r$ as equivalent to variations in the resonator spring constant $k_r$. In this paper we investigate the range $0.03 \lesssim \Omega \lesssim 30$. All results for the following parametric studies are obtained through numerical analysis conducted using MathWorks MATLAB.

\begin{figure*}
 \includegraphics[width=2\columnwidth]{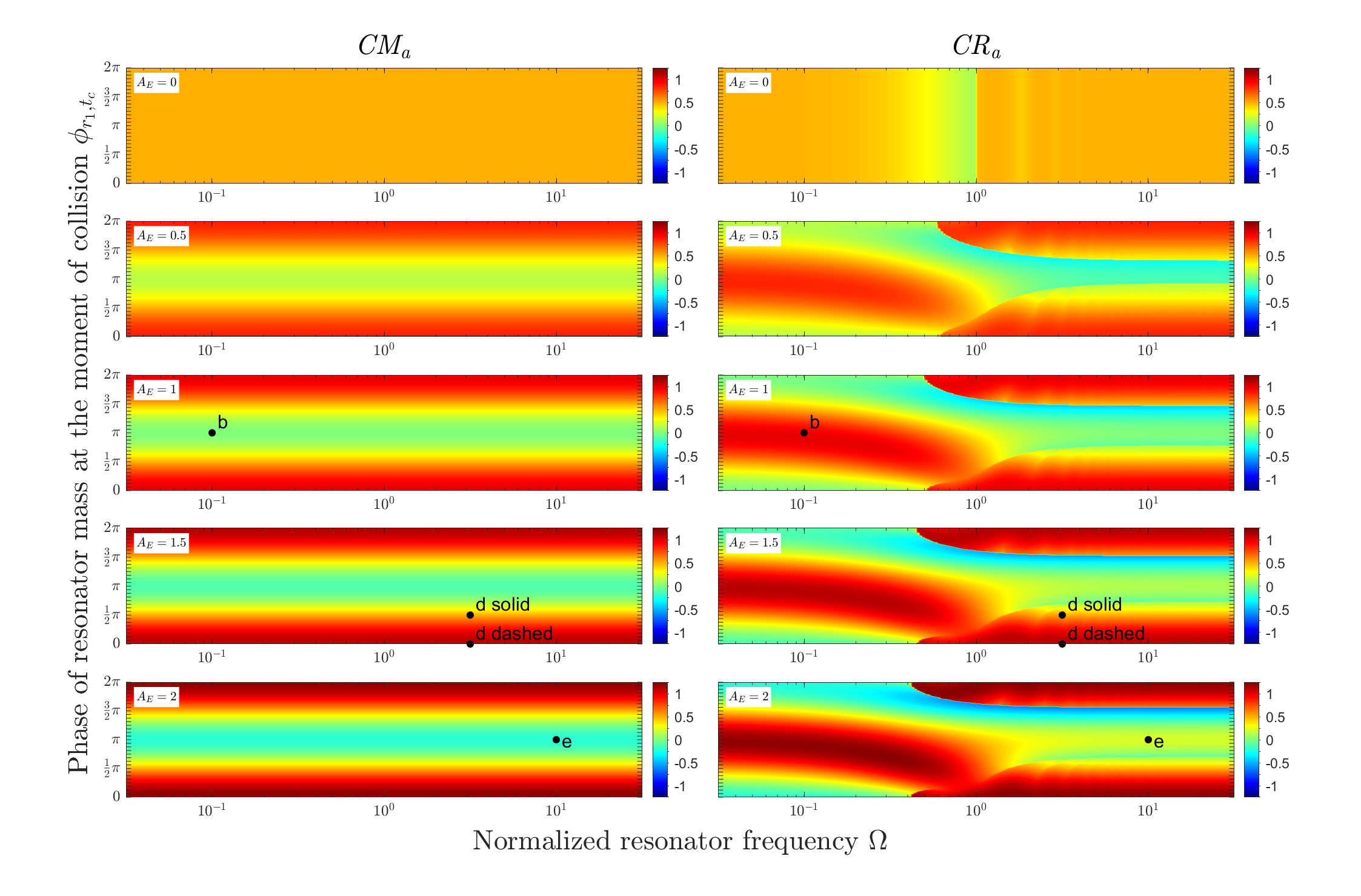}
 \caption{Density plots for the effective coefficients of motion, $CM_a$ (left panels), and restitution, $CR_a$ (right panels), for the chosen mass-in-mass resonator system. The vertical axes show the phase of the resonator at the moment of collision, and the horizontal axes show the normalized frequency ratio $\Omega$. Each plot represents the results for different initial conditions of the left-hand mass-in-mass system at the moment of collision. The order from top to bottom in each case corresponds to increasing total energy of the resonator. The labeled data points correspond to selected parameters, as shown in the panels of Fig.~\ref{Figure5}.}\label{Figure2}
 \end{figure*}

Figure~\ref{Figure2} shows density plots for the effective coefficients of motion $CM_a$ (left panels) and restitution $CR_a$ (right panels), with the vertical axes showing the phase of the resonator at the moment of collision and the horizontal axes showing the normalized frequency ratio $\Omega = \omega_r / \omega_c$, for different resonator initial conditions.

Conditions for a stiff sphere are implemented using a steel shell with a radius of $R = 10$ mm and shell thickness $0.1R$, leading to $\omega_c / \omega_g = 4.7 \times 10^4$. The initial conditions of the internal resonator of the (left-hand) mass-in-mass system 1 at the moment of collision are described in terms of $E_n$, the total energy (including both kinetic and potential energy) of the resonator relative to the kinetic energy of an equivalent non-resonant sphere of mass $m$ (which increases down each column, with its value indicated in each panel), and $\phi_{r_1, t_c}$, the phase of the internal resonant mass relative to its undisturbed location (plotted on the vertical axis of each panel). Further,  $x_{r_1,t_c}/\dot{x}_{1,t_c}=\sqrt{E_n/\omega_r^2}\sin{\phi_{r_1,t_c}}$ and $\dot{x}_{r_1,t_c}/\dot{x}_{1,t_c}=\sqrt{E_n}\cos{\phi_{r_1,t_c}}$. A phase of 0 or $\pi$ corresponds to no potential energy stored in the resonator, and a phase of $\pi/2$ or $3\pi/2$ corresponds to no kinetic energy stored in the resonator. Green represents zero on the color bar, with positive values indicated by warmer colors and negative values by cooler colors.

The top panel in Fig.~\ref{Figure2} represents the case in which no energy is stored in the resonator of the mass-in-mass system 1 at the moment of impact. As expected from Eqs.~(\ref{eq:CM_a_effective}), $CM_a$ is fixed at 0.5, which is equivalent to the non-resonant case described earlier with $m + m_r = 2m$. This case also eliminates the effect of the phase parameter of the resonator on the effective coefficient of restitution, $CR_a$. The effective coefficient of restitution is still a function of the resonator frequency. It exhibits a resonant-like effect, showing a dip to zero as the resonator frequency approaches the compression frequency from below. The asymptotic values correspond to the non-resonant case.

By introducing initial conditions with non-zero resonator energy of the mass-in-mass system 1, a much larger range and combination of effective coefficients can be achieved. For both effective parameters, the sign of the resonator's velocity has the strongest effect on the relative change of their values.

$CR_a$ in Eqs.~(\ref{eq:v1_and_v2_CR_effective}) is directly equivalent to the conventional coefficient of restitution in Eqs.~(\ref{eq:v1_and_v2_CR_conv_identical}). In the conventional case, when there is no attractive force between the entities during the collision and no process that causes the particles to stick together, this coefficient represents how well kinetic energy is conserved during the collision. It ranges from 0 for a completely inelastic collision to 1 for a completely elastic collision. A value of 0 indicates that the final momentum is equally distributed, and the entities couple together, whereas a value of 1 implies that all of the momentum is transferred to the second entity, provided they have equal mass.

Figure~\ref{Figure2} shows that the effective coefficient of restitution $CR_a$ can exceed 1, indicating that kinetic energy is being added to the spherical shells. The source of this kinetic energy is the energy stored in the resonator of mass-in-mass system 1 at the moment of collision. For resonator frequencies below the compression frequency during contact (i.e., $\Omega < 1$), $CR_a$ tends to increase when the resonator's phase is $\pi$ (negative resonator velocity) and decrease when the phase is 0 or $2\pi$ (positive resonator velocity). This trend reverses for resonator frequencies above the compression frequency (i.e., $\Omega > 1$).

Figure~\ref{Figure2} also shows that the coefficient of restitution $CR_a$ can take negative values. However, unless the average displacement of mass-in-mass system 2 in the external potential is equal to or greater than the (positive) displacement of system 1 after impact, the entities will collide again shortly after the initial collision. The exact timing of this subsequent collision depends on the phase of the resonant oscillatory motion superimposed on the motion in the general external potential. As indicated by Eqs.~(\ref{eq:pendulum_after_collision_inst}) and (\ref{eq:x_effective_a}), negative values of $\mathit{CR}_a$ will quickly lead to fragmentation of the general oscillatory behavior in the external potential.

The effective coefficient of motion $CM_a$ in Eqs.~(\ref{eq:v1_and_v2_CR_effective}) is equivalent to the fixed unity term in Eqs.~(\ref{eq:v1_and_v2_CR_conv_identical}), which in the conventional case of identical simple spheres results from the conservation of momentum. This new term therefore arises from the effective momentum conservation of the spherical shells. Unlike in the conventional case, it can vary since the outer spherical shells on their own are not a closed system, i.e., $CM_a$ is governed by the total momentum of the spherical shells post-collision relative to the value of the momentum pre-collision. Figure~\ref{Figure2} and Eqs.~(\ref{eq:CM_a_effective}) show that $CM_a$ is not a function of the resonator frequency $\omega_r$, but rather is simply determined by the velocity of the resonant inner mass at the moment of collision. A positive value is associated with an effective addition of momentum to the system of spherical shells. In contrast, a negative value is associated with a reduction in the momentum of the system of spherical shells; when $\dot{x}_{r,t_c}<-\dot{x}_{1, t_c}$, the momentum of the system of spherical shells is negative (i.e., directed in the opposite direction) relative to their momentum at the moment of collision. 

\begin{figure}
 \includegraphics[width=1\columnwidth]{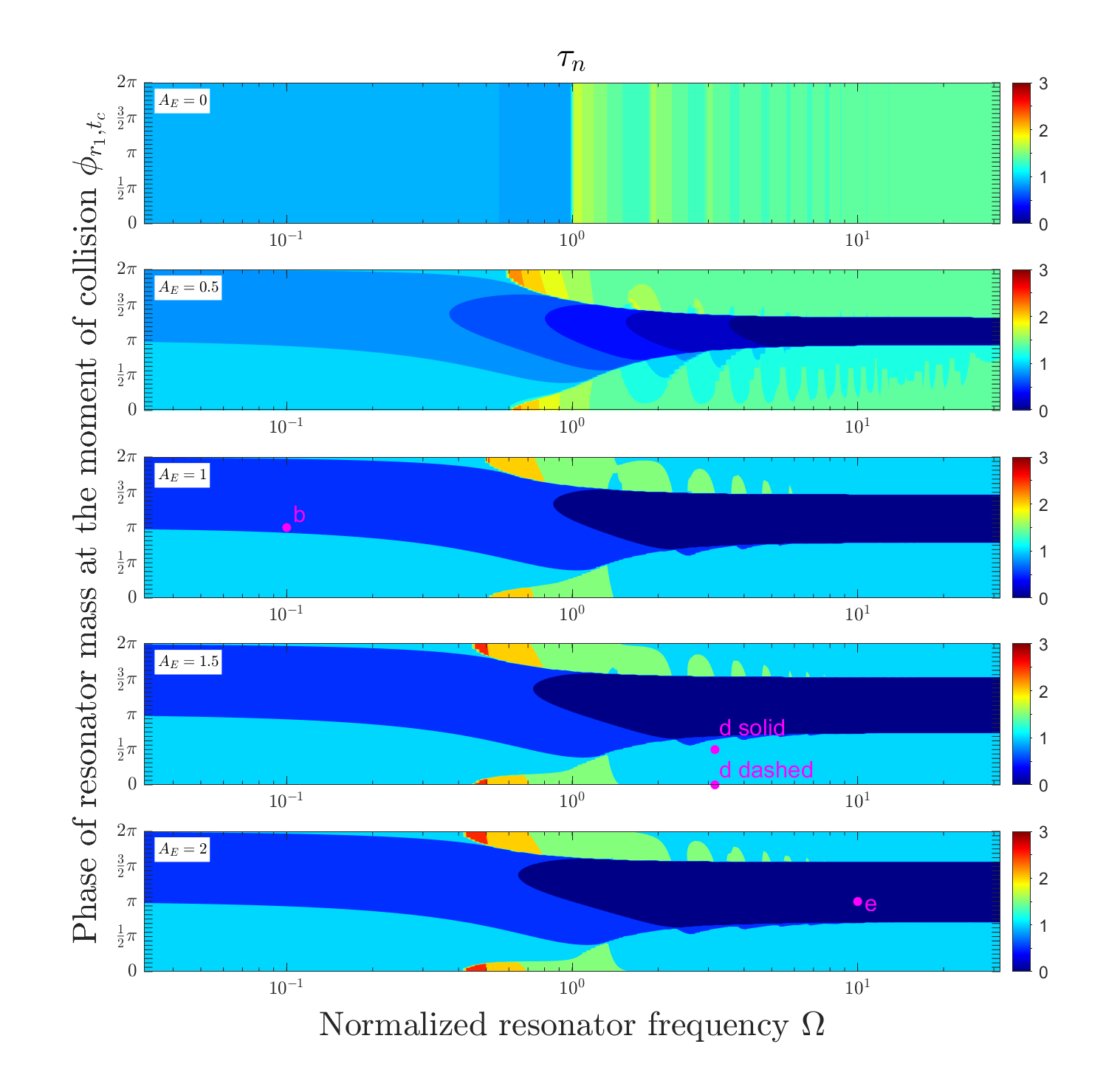}
 \caption{Density plots of $\tau_n$, the contact duration of the chosen mass-in-mass resonator system, normalized to the contact duration for an equivalent shell containing no internal mass, are shown for the selected mass-in-mass resonator system. The vertical axes represent the phase of the resonator at the moment of collision, whereas the horizontal axes show the normalized frequency ratio $\Omega$. The order from top to bottom in each case corresponds to increasing total energy of the resonator. The labeled data points correspond to selected parameters, as shown in the panels of Fig.~\ref{Figure5}.}\label{Figure3}
 \end{figure}

Figure~\ref{Figure3} shows equivalent density plots for $\tau_n$, the contact duration of the chosen mass-in-mass resonator system, normalized to the contact duration for an equivalent shell containing no internal mass. The vertical axis shows the phase of the resonator at the moment of collision, whereas the horizontal axis shows the normalized frequency ratio $\Omega$, for different resonator initial conditions. The contact duration was determined from numerical simulations of the collision using Eqs.~(\ref{eq:contact}).

The general effect of the local resonance is to reduce the contact time when the resonator frequency is above the compression frequency and the resonator velocity is negative, i.e., when the phase of the internal mass lies in the interval $\phi_{r_1,t_c} = (-\pi/2, \pi/2)$. In this case, the initial motion of the internal resonator mass pulls the spherical shell 1 back from the initially stationary spherical shell 2. The contact duration remains relatively unchanged for $\Omega < 0.1$, as the oscillation of the internal mass is slower compared to the compression frequency, resulting in a smaller effect on the compression time.

The strongest effect occurs when $\phi_{r_1,t_c} \approx 0 \text{ or } \pi$, corresponding to positive resonator velocities on impact, and when the resonator frequency is just below the compression frequency, with $\Omega \approx$ 0.4 to 1. In this case, the internal resonator mass pulls the spherical shell 1 towards shell 2. Since the resonator frequency is lower than the compression frequency, the resonator velocity changes more slowly and continues to pull shell 1 towards shell 2, prolonging the contact.

\begin{figure*}
 \includegraphics[width=2\columnwidth]{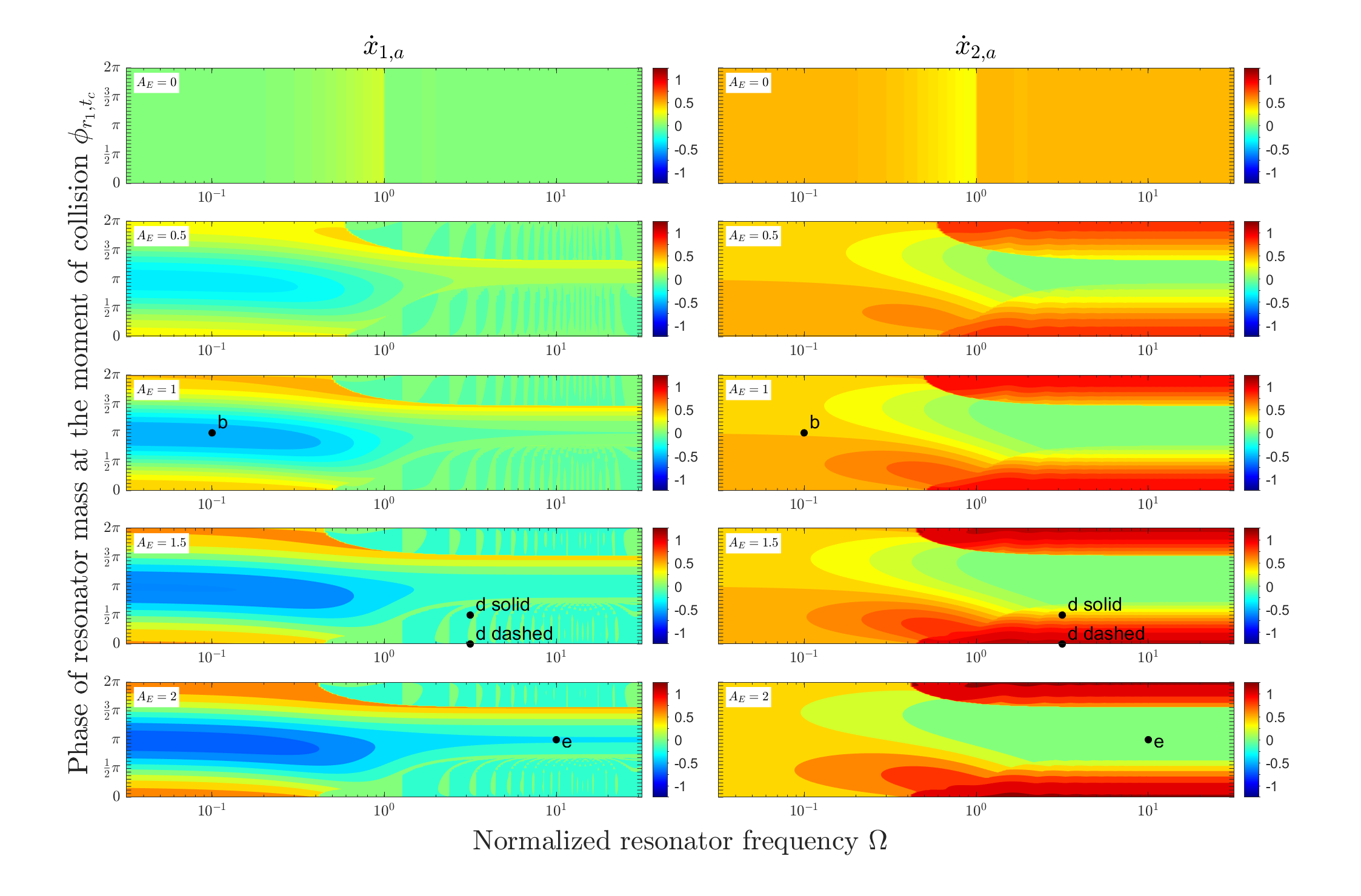}
 \caption{Density plots of the initial velocity of spherical shell 1 (left column) and shell 2 (right column) of the mass-in-mass system at the moment of separation, normalized to the velocity of spherical shell 1 at the initial time of collision. The vertical axes show the phase of the resonator at the moment of collision, whereas the horizontal axes show the normalized frequency ratio $\Omega$. The order from top to bottom in each case corresponds to increasing total energy of the resonator. The labeled data points correspond to selected parameters, as shown in the panels of Fig.~\ref{Figure5}.}\label{Figure4}
 \end{figure*}

Figure~\ref{Figure4} shows density plots similar to those in Fig.~\ref{Figure2}, but for the initial velocities of each spherical shell in the external potential after impact (shell 1 in the left column and shell 2 in the right column), normalized to the velocity of shell 1 at the moment of collision. This is equivalent to Eqs.~(\ref{eq:x_effective_a}) normalized by $\dot{x}_{1,t_c}$. While the effective parameters in Fig.~\ref{Figure2} provide insight into the overall properties of the mass-in-mass systems post-collision, the initial velocity parameters in Fig.~\ref{Figure4} highlight the response of each spherical shell in the external potential. This demonstrates that the mass-in-mass resonator system can exhibit a variety of responses in an external potential depending on the selected resonator properties.

\subsection{Simulated motion of the dynamics of a particular system}

\begin{figure*}
 \includegraphics[width=0.73\columnwidth]{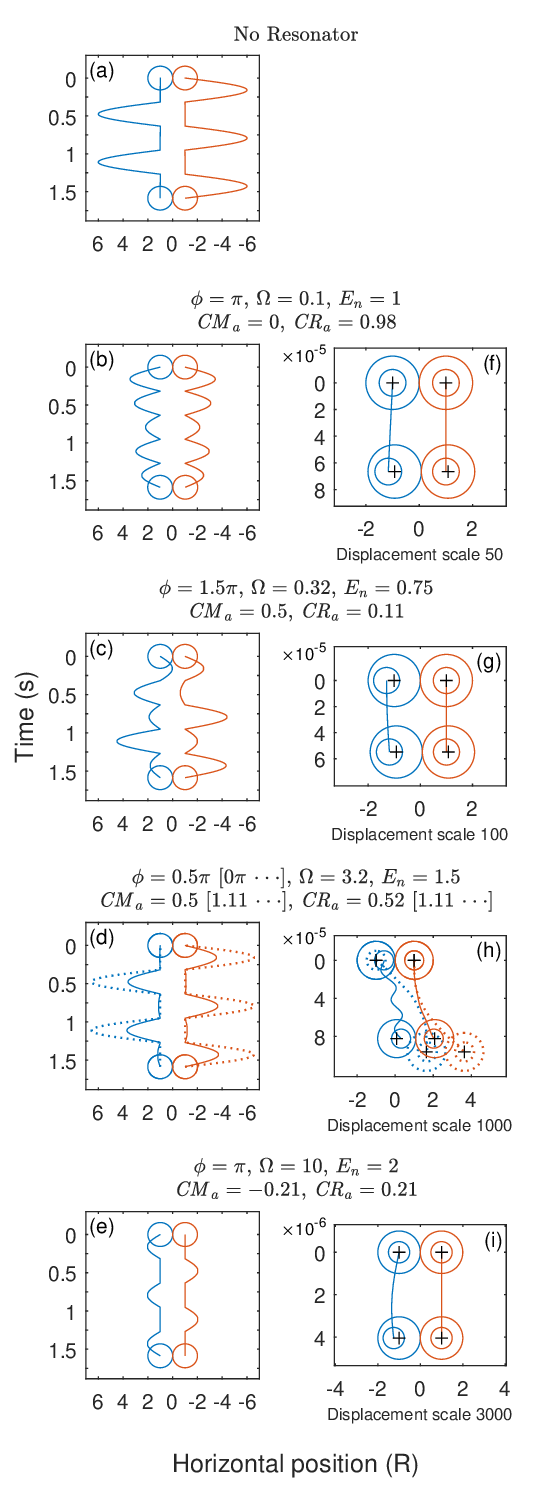}
 \caption{Time-domain displacement plots showing the motion of the two spherical shells for selected resonator parameters and initial conditions corresponding to a range of values of the effective coefficients of motion, $\it{CM}_{a}$, and restitution, $\it{CR}_{a}$. Sphere 2 and its resonator are initially stationary at the equilibrium point (with zero kinetic and potential energy) at the time of collision (time = 0~s). The left-hand panels (a)-(e) show the motion of the outer shells over the first six collisions. Panel (a) represents the time response for the case of solid spheres with no internal resonator, whereas panels (b)-(e) show the time response for different parameters and initial conditions, as indicated in each panel. The right-hand panels (f)-(i) display the displacement of both the outer and inner shells (scaled as indicated below each panel) during the initial collision for each sequence shown in the corresponding left panel. The black crosses indicate the centers of each outer sphere at the start and end times.}\label{Figure5}
 \end{figure*}

We now present the real-time dynamics of the system, which are crucial for understanding its behavior. To illustrate how different values of the effective coefficients influence post-collision motion, the left-hand panels (a)–(e) in Fig.~\ref{Figure5} show the simulated trajectories of the spherical shells over six successive collisions in an external gravitational potential. Time is plotted on the vertical axis, increasing downward from zero, allowing the relative displacement of the spherical shells in the horizontal direction to be observed over time. The right-hand panels (f)–(i) in Fig.~\ref{Figure5} depict the motion, during the first collision of the spherical shells, of the outer shells and internal masses for the motion depicted in the corresponding left-hand panels. Since the actual displacement during the collision is much smaller than the dimensions of the spheres, the displacements in these panels have been scaled by the values indicated below each figure to better visualize the relative motion. These results clearly demonstrate that even very small displacements of the internal masses—typically at least three orders of magnitude smaller than those of the outer shells—can significantly influence the post-collision motion of the shells. The displacements are obtained from numerical simulations (MathWorks Simulink) of the full equations of motion, incorporating both the collision dynamics from Eqs.~(\ref{eq:contact}) and the oscillations in the external potential described by Eqs.~(\ref{eq:pendulum}), for a range of representative values and combinations of the effective coefficients. Animations of the system's time response for each scenario are available in the Supplemental Material. Figure~\ref{Figure5}(a) represents the conventional case of two simple spheres, each with mass $m$.

Figure~\ref{Figure5}(b) illustrates the response of the outer spherical shells for a chosen normalized resonator frequency $\Omega=0.1$, phase $\phi=\pi$ and normalized initial energy $E_n = 1$, corresponding to $\it{CM}_{a} =$~0 and $\it{CR}_{a} =$~0.98. The condition $\it{CM}_a = 0$ implies that momentum is not conserved for the spherical shells, causing them to recoil with equal and opposite velocities after the collision. The coefficient $\it{CR}_{a}$ determines the fraction of energy retained by the spherical shells post-collision, thereby setting the magnitude of their velocities. With $\it{CR}_{a} = $~0.98, the collision is nearly elastic, meaning that the velocity magnitude of each shell immediately after impact is approximately half of shell 1’s velocity at the moment of collision.

The peak amplitudes of displacement in the external potential are also roughly equal and opposite, as they result from integrating the first two velocity terms in Eqs.~(\ref{eq:pendulum_after_collision}) when their phase is $\pi/2$. In conventional collisions between simple spheres, such opposite velocities occur only if the second shell is significantly more massive than the first. Since these effective parameters apply specifically to the first collision, subsequent collisions may exhibit variations in the oscillatory dynamics, but the overall motion remains approximately equal and opposite between the two outer shells.

Figure~\ref{Figure5}(f) further reveals that at the end of the first collision, the internal mass of shell 1 moves leftward due to its initial negative velocity. This motion pulls shell 1 back in the opposite direction of its initial trajectory at impact. In contrast, the internal mass of shell 2 remains relatively stationary, allowing shell 2 to move as expected for a conventional post-collision response, albeit with a smaller velocity, since part of the pre-impact energy is retained by shell 1.

Figure~\ref{Figure5}(c) shows the response for $\Omega=0.32$, $\phi=1.5\pi$ and $E_n = 0.75$, chosen to correspond to $\it{CM}_a =$~0.5 and $\it{CR}_a =$~0.11. The case $\it{CR}_a = 0$ corresponds to a perfectly inelastic collision, in which the spherical shells appear to couple together after impact and move forward in near-unison. However, as mentioned above, the representative value of $\it{CR}_a =$~0.11 is selected for the simulation. This prevents the shells from colliding shortly after separation due to the oscillation in $x_{i,r}$, which results from the motion of the internal resonators\textemdash a motion superimposed on the average displacement variation $x_{i,a}$ of the shells. The value $\it{CM}_a =$~0.5 implies that the spherical shells possess approximately half of the total momentum of shell 1 before impact. This leads to a peak displacement for each shell that is approximately $1/4$ of the displacement obtained in the conventional case with simple spheres of mass $m$.  

Figure~\ref{Figure5}(g) shows that the initial peak negative displacement of the internal mass of shell 1 remains relatively unchanged during the collision. The resulting positively directed motion of the internal mass immediately after the collision acts to pull the outer shell to the right, leading in turn to its positively directed motion. The internal mass of shell 2 remains relatively stationary, resulting in a motion similar to that observed in Figs.~\ref{Figure5}(b) and (f).  

Figure~\ref{Figure5}(d) shows the response for $\Omega=3.2$ and $E_n = 1.5$, but with two different cases of resonator initial phase that yield $\it{CR}_a \sim \it{CM}_a$. The overall dynamics closely follow those of the conventional case for a fully elastic collision; that is, shell 1 remains stationary after the first collision. However, the magnitude of the displacement of shell 2 for $\it{CM}_a = \it{CR}_a =$~1.1, when $\phi = 0\pi$ (dashed line), is $10\%$ greater than in the conventional case, whereas the values of $\it{CM}_a = \it{CR}_a =$~0.5 (solid line), when $\phi = 0.5\pi$, correspond to $0.5$ times that of the conventional case.

Such a situation can only occur in the conventional case when the two solid spheres have different masses ($m_1/m_2 \neq 0$). Figure~\ref{Figure5}(h) shows that the displacement of the shells during the first collision is significant for both sets of effective parameters. This is caused by a combination of the internal resonant frequency being greater than the compression frequency of conventional solid spheres and the longer contact time. This leads to enhanced energy transfer between the internal mass and outer shells during the collision.  

The positive momentum of the internal mass for the case corresponding to $\it{CM}_a = \it{CR}_a =$~1.1 (dashed line) is partially transferred to outer shell 2, leading to the enhanced displacement observed in the left-hand panel. In contrast, the peak positive displacement of internal mass 1 at the time of collision for $\it{CM}_a = \it{CR}_a =$ 0.5 means that the initial motion of the internal mass is negative during the collision, resulting in less net momentum transfer. 

Figure~\ref{Figure5}(e) shows the response of the shells for $\Omega=10$, $\phi=\pi$ and $E_n = 2$, chosen to correspond to $\it{CM}_{a} <$~0. For the case in which $\it{CR}_{r} \approx -\it{CM}_{a}$, the overall dynamics resemble those of the conventional case of a solid sphere colliding with a rigid surface (approximated by $m_2 \gg m_1$, where 1 and 2 refer to the two entities). That is, shell 1 bounces back while shell 2 remains stationary. When this occurs, zero net momentum is transferred on average to shell 2, so that it remains stationary and effectively acts like a fixed surface. In contrast, a sufficient amount of negative momentum, relative to the momentum at the time of collision, is transferred from the internal mass to shell 1, causing it to rebound. For this to occur, the phase of the internal resonator in shell 1 at the moment of collision should be approximately $\pi$. This corresponds to a relatively large and negative internal mass velocity, meaning that the internal resonator acts to push shell 1 back during the collision. Figure~\ref{Figure5}(i) also indicates a significantly reduced contact time, which leads to less momentum transfer to shell 2.

\section{Conclusions}

In conclusion, we have analyzed the collision dynamics of a metamaterial two-sphere Newton's cradle system and derived theoretical expressions for the effective coefficients of restitution and motion. The analysis focused on the case in which the local resonance can be considered strong, that is, where it has a much higher resonant frequency than the independent oscillations of the two mass-in-mass spheres in the external gravitational potential.

The effective parameters are presented in a form that is also convenient for application to a general periodic lattice. We show that using a two-sphere mass-in-mass system in a Newton's cradle geometry results in more complex behavior than in the conventional case, depending on the initial conditions of the internal resonator mass at the moment of collision. Our results demonstrate that the effective coefficient of motion can be negative, causing the momentum of the outer spherical shells to be reversed by the inner-mass local resonance properties. The effective coefficient of restitution exhibits a range of values, providing a much richer spectrum of dynamic behavior compared to the conventional two-sphere Newton's cradle. We also show that, in general, the magnitude ranges of both effective parameters increase with increasing resonator energy, while the contact duration of the shells decreases. The sign of the resonator velocity at the moment of collision is shown to have a significant effect on the subsequent dynamics.

Our derived effective parameters suggest the existence of exotic properties for an equivalent periodic lattice of mass-in-mass unit cells, which may have practical applications. These results are particularly relevant for the design of effective dynamical properties in metamaterials based on local resonant characteristics, as they demonstrate how overall momentum and energy conservation of the effective properties can be tailored through the initial local resonator conditions. Further work is needed to extend these results to include both larger amplitude oscillations and frictional effects. Finally, the continuation of this research should lead to the construction and investigation of new classes of ``collision'' metamaterial structures.

\begin{acknowledgments}
SAP was supported by the UK Engineering and Physical Sciences Research Council through grant EP/T028661/1. OBW acknowledges Grants-in-Aid for Scientific Research from the Ministry of Education, Culture, Sports, Science and Technology (MEXT).

\end{acknowledgments}


\bibliography{Metamaterialcollisionreferences}

\end{document}


\section*{SUPPLEMENTAL MATERIAL}

\title{Metamaterial two-sphere Newton's Cradle}


\author{Simon A. Pope}
\email[]{s.a.pope@sheffield.ac.uk}
\affiliation{Department of Automatic Control and Systems Engineering, University of Sheffield, Mappin Street, Sheffield, S1 3JD, UK.}

\author{Oliver B. Wright}
 \affiliation{Hokkaido University, Sapporo, 060-0808, Japan}
 \affiliation{Graduate School of Engineering, Osaka University,
   2-1 Yamadaoka, Suita, Osaka, 565-0871, Japan}


\date{\today}
\maketitle

\begin{section}{Derivation of the pendulum dynamics of the two-sphere mass-in-mass Newton's cradle}
For the system in Fig.~1 (main text) and reproduced in Fig.~\ref{Figure1}, the displacement vector for the center of mass of outer shell $i$ and its inner resonant mass is given in Cartesian coordinates by
$$
\begin{aligned}
& \left[\begin{array}{c}x_i \\ y_i \end{array}\right]
=\left[\begin{array}{c}
L \sin \theta_i \\
L-L \cos \theta_i
\end{array}\right], \\
& \left[\begin{array}{c}x_{r_i} \\ y_{r_i} \end{array}\right]
=\left[\begin{array}{c}x_i \\ y_i \end{array}\right]
+\left[\begin{array}{c}
\Delta x_{r_i} \cos \theta_i \\
\Delta x_{r_i} \sin \theta_i
\end{array}\right].
\end{aligned}
$$
Here, $\theta_i$ is the angular displacement of the pendulum, $\Delta x_{r_i}$ is the extension of the internal spring (i.e., the displacement of the internal mass relative to the center of the sphere), and $x_i$, $y_i$ are the horizontal and vertical displacements, respectively, of the outer shell relative to an origin at the undisturbed position of the sphere/mass system. The coordinates $x_{r_i}$, $y_{r_i}$ describe the position of the inner resonant mass.
 \begin{figure}[!hb]
  \includegraphics{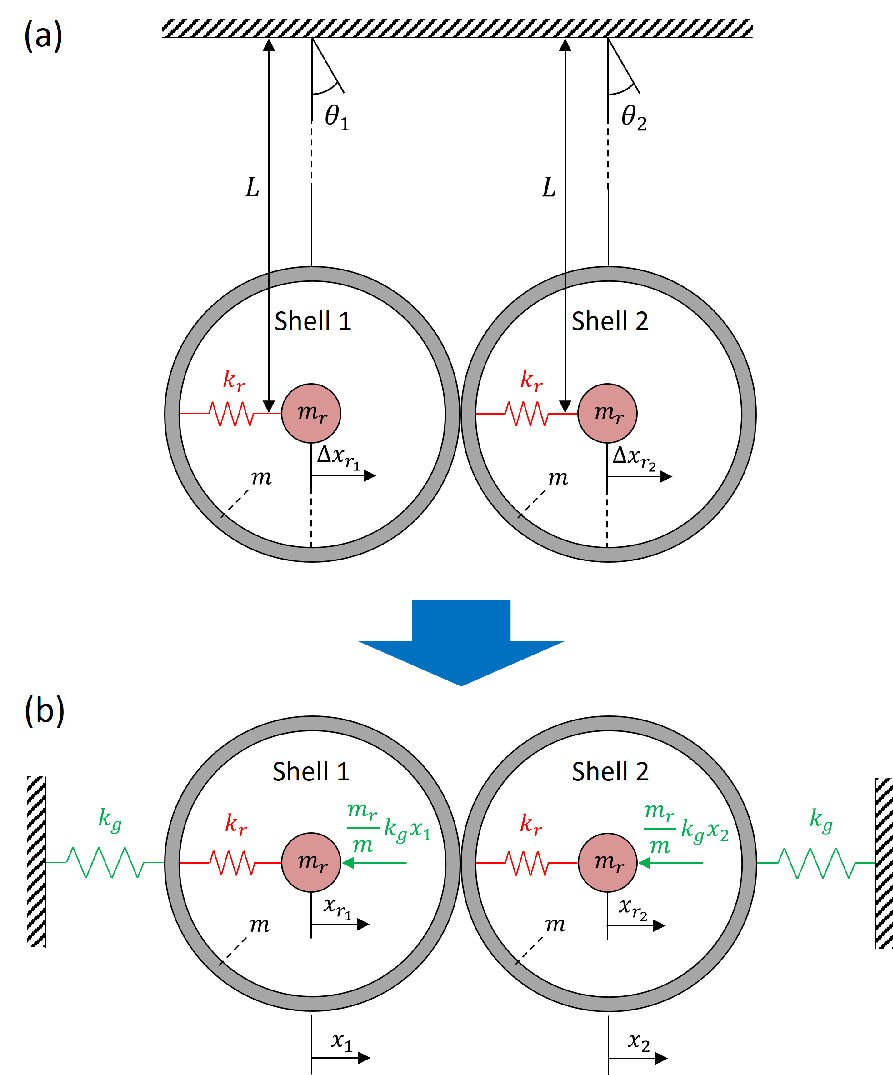}
 \caption{(a) The two-sphere mass-in-mass colliding pendulum system (a modified Newton's cradle), along with (b) the equivalent approximate pure mass-and-spring model applicable for small displacements and stiff spherical shells, which exhibits one-dimensional motion. The chosen model for the dynamics during a collision, which depends on the collision stiffness $k_c$ in a linear approximation, is not shown.} \label{Figure1}
 \end{figure}

This leads leads to velocities and accelerations in the Cartesian frame:
$$
\begin{aligned}
& \left[\begin{array}{c}\dot{x}_i \\ \dot{y}_i \end{array}\right]=\left[\begin{array}{c}
L\dot{\theta_i} \cos \theta_i \\
L \dot{\theta_i} \sin \theta_i
\end{array}\right], \\ 
& \left[\begin{array}{c}\dot{x}_{r_i} \\ \dot{y}_{r_i} \end{array}\right]
= \left[\begin{array}{c}\dot{x}_i \\ \dot{y}_i \end{array}\right]
+ \Delta \dot{x}_{r_i} \left[\begin{array}{c}
\cos \theta_i \\
\sin \theta_i
\end{array}\right]
+\Delta x_{r_i} \dot{\theta} \left[\begin{array}{c}
-\sin \theta_i \\
\cos \theta_i
\end{array}\right].
\end{aligned}
$$
and
$$
\begin{aligned}
& \left[\begin{array}{c}\ddot{x}_i \\ \ddot{y}_i \end{array}\right]=\left[\begin{array}{c}
L\left(\ddot{\theta_i} \cos \theta_i  - \dot{\theta_i}^2 \sin \theta_i\right) \\
L \left(\ddot{\theta_i} \sin \theta_i + \dot{\theta_i}^2 \cos \theta_i \right)
\end{array}\right], \\
& \left[\begin{array}{c}\ddot{x}_{r_i} \\ \ddot{y}_{r_i} \end{array}\right]
= \left[\begin{array}{c}\ddot{x}_i \\ \ddot{y}_i \end{array}\right]
+\Delta \ddot{x}_{r_i} \left[\begin{array}{c}
\cos \theta_i \\
\sin \theta_i
\end{array}\right]
+2\Delta \dot{x}_{r_i} \dot{\theta} \left[\begin{array}{c}
-\sin \theta_i \\
\cos \theta_i
\end{array}\right]
+\Delta {x}_{r_i} \ddot{\theta}_i \left[\begin{array}{c}
-\sin \theta_i \\
\cos \theta_i
\end{array}\right]
+\Delta x_{r_i} \dot{\theta}^2 \left[\begin{array}{c}
-\cos \theta_i \\
-\sin \theta_i
\end{array}\right]. 
\end{aligned}
$$

The kinetic energy of the outer sphere is given by
$$
\begin{aligned}
T_m = \frac{1}{2}m\left(L\dot{\theta}_i\right)^2,
\end{aligned}
$$
whereas the kinetic energy of the resonant mass is
$$
\begin{aligned}
T_{m_r} & = \frac{1}{2}m_r \left(\dot{x}_{r_i}^2+\dot{y}_{r_i}^2\right), \\
& = \frac{1}{2}m_r\left( \left(L \dot{\theta_i} \cos \theta_i+\left(\Delta \dot{x}_{r_i} \cos \theta_i-\Delta x_{r_i} \dot{\theta_i} \sin \theta_i\right)\right)^{2}+\left(L\dot{\theta_i} \sin \theta_i+\left(\Delta \dot{x}_{r_i} \sin \theta_i + \Delta x_{r_i} \dot{\theta_i} \cos \theta_i\right)\right)^{2}\right), \\
& =\frac{1}{2} m_r\left(L^{2} \dot{\theta_i}^{2}+\Delta \dot{x}_{r_i}^{2}+\dot{\theta_i}^{2} \Delta x_{r_i}^{2}+2 L \dot{\theta_i} \Delta \dot{x}_{r_i}\right).
\end{aligned}
$$
The potential energy of the outer sphere taken form the undisturbed position ($\theta_i=0$) as the reference is given
$$
\begin{aligned}
V_m = mg\left(L-L\cos\theta_i\right).
\end{aligned}
$$
The potential energy of the resonant mass relative to its undisturbed position ($\theta_i = 0$ and $\Delta x_{r_i} = 0$) is given by
$$
\begin{aligned}
V_{m_r }= m_rg\left(L-L\cos\theta_i+\Delta x_{r_i} \sin \theta_i\right),
\end{aligned}
$$
whereas the potential energy of the spring is 
$$
\begin{aligned}
V_{k_r} = \frac{1}{2}k_r\Delta x_{r_i}^2.
\end{aligned}
$$
The full expression for the kinetic and and potential energy of the mass-in-mass pendulum is given by
$$
\begin{aligned}
& T=T_{m} + T_{m_r}, \\
& V=V_{m}+V_{m r}+V_{k_{r}}.
\end{aligned}
$$

The Lagrangian is defined as $L=T-V$, and the equations of motion for an unforced system are calculated for each independent coordinate $q_i$ from the following:
$$
\begin{aligned}
& \frac{\partial}{\partial t}\left(\frac{\partial L}{\partial \dot{q}_{i}}\right)-\frac{\partial L}{\partial q_{i}}=0.
\end{aligned}
$$
The required derivatives of the total kinetic and potential energy functions with respect to the two independent coordinates are given by
$$
\begin{aligned}
& \frac{\partial T}{\partial \dot{\theta_i}}=m L^{2} \dot{\theta_i}+m_r\left(L^{2} \dot{\theta_i}+\dot{\theta_i} \Delta x_{r_i}^{2}+L \Delta\dot{x}_{r_i}\right), \\
& \frac{\partial V}{\partial \dot{\theta_i}}=0, \\
& \frac{\partial T}{\partial \Delta \dot{x}_{r_i}}=m_{r}\left(\Delta \dot{x}_{r_i}+L \dot{\theta_i}\right), \\
& \frac{\partial V}{\partial \Delta \dot{x}_{r_i}}=0, \\
& \frac{\partial T}{\partial \theta_i}=0, \\
& \frac{\partial V}{\partial \theta_i}=m g L \sin \theta_i+m_r g\left( L \sin \theta_i+\Delta x_{r_i} \cos \theta_i\right), \\
& \frac{\partial T}{\partial \Delta {x_{r_i}}}=m_{r} \dot{\theta_i}^{2} \Delta x_{r_i}, \\
& \frac{\partial V}{\partial \Delta {x_{r_i}}}=k_{r} \Delta {x_{r_i}}+m_{r} g \sin \theta_i.
\end{aligned}
$$
The equation of motion for the outer sphere is
\begin{eqnarray}
& \frac{\partial}{\partial t}\left(\frac{\partial L}{\partial \dot{\theta_i}}\right)-\frac{\partial L}{\partial \theta_i}=0, \nonumber \\
& m L \ddot{\theta}_i + m_r\left(L^2\ddot{\theta}_i+\ddot{\theta}_i\Delta x_{r_i}^2+2\dot{\theta}_i\Delta x_{r_i}\Delta \dot{x}_{r_i}+L\Delta \ddot{x}_{r_i}\right)+m g L \sin \theta_i+m_{r} g\left(L \sin \theta_i+\Delta x_{r_i}\cos \theta_i\right)=0. \label{suppeq_m}
\end{eqnarray}
The equation of motion for the inner mass is
\begin{eqnarray}
& \frac{\partial}{\partial t}\left(\frac{\partial L}{\partial \Delta \dot{x}_{r_i}}\right)-\frac{\partial L}{\partial \Delta x_{r_i}}=0, \nonumber \\
& m_{r}\left(\Delta \ddot{x}_{r_i}+L \ddot{\theta_i}\right)-m_{r} \dot{\theta_i}^2 \Delta x_{r_i}+m_r g \sin \theta_i+k_{r} \Delta x_{r_i}=0. \label{suppeq_mr}
\end{eqnarray}
Eq.~(\ref{suppeq_mr}) can be written as
\begin{equation}
m_{r}\left(L \Delta \ddot{x}_{r_i}+L^{2} \ddot{\theta_i}\right)+m_{r} g L \sin \theta_i=m_{r} L \dot{\theta_i}^{2} \Delta x_{r_i}-L k_{r} \Delta x_{r_i}, \nonumber
\end{equation}
which on substitution into Eq.~(\ref{suppeq_m}) gives an alternative form for the equation of motion of the outer sphere:
\begin{equation}
m L^{2} \ddot{\theta_i}+m_{r}\left(\ddot{\theta_i} \Delta x_{r_i}^{2}+\dot{\theta_i} \Delta x_{r_i} \Delta \dot{x}_{r_i} + L \dot{\theta_i}^{2} \Delta x_{r_i}\right)+m g L \sin \theta_i +m_{r} g \Delta x_{r_i} \cos \theta_i-L k_{r} \Delta x_{r_i}=0. \label{suppeq_mnew}
\end{equation}

So that the system is restricted to normal collisions along the center of mass of each sphere, we consider only small angles of displacement of the pendulums. The equations of motion Eqs.~(\ref{suppeq_mnew}) and (\ref{suppeq_mr}) for the sphere and internal resonant mass on application of the small-angle approximation become
\begin{eqnarray}
& m L^{2} \ddot{\theta_i}+m_{r}\left(\ddot{\theta_i} \Delta x_{r_i}^{2}+\dot{\theta_i} \Delta x_{r_i} \Delta \dot{x}_{r_i}+L \dot{\theta_i}^{2} \Delta x_{r_i}\right)+m g L \theta_i +m_{r} g \Delta x_{r_i} -L k_{r} \Delta x_{r_i}=0, \label{suppeq_mnew_small} \\
& m_{r}\left(\Delta \ddot{x}_{r_i}+L \ddot{\theta_i}\right)-m_{r} \dot{\theta_i}^2 \Delta x_{r_i}+m_r g \theta_i+k_{r} \Delta x_{r_i}=0. \label{suppeq_mr_small}
\end{eqnarray}
Under such conditions the small-angle approximation also leads to displacement vectors
$$
\begin{aligned}
& \left[\begin{array}{c}x_i \\ y_i \end{array}\right]=\left[\begin{array}{c}
L \theta_i \\
0
\end{array}\right], \\
& \left[\begin{array}{c}x_{r_i} \\ y_{r_i} \end{array}\right]
=\left[\begin{array}{c}x_i \\ y_i \end{array}\right]
+\left[\begin{array}{c}
\Delta x_{r_i} \\
\Delta x_{r_i} \theta_i
\end{array}\right]
=\left[\begin{array}{c}x_i \\ 0 \end{array}\right]
+\left[\begin{array}{c}
\Delta x_{r_i} \\
\frac{\Delta x_{r_i}}{L} x_i
\end{array}\right].
\end{aligned}
$$
The small-angle approximation only leads to a single-dimensional system if the extension of the spring $\Delta x_{r_i}$ is much smaller than the length of the pendulum, i.e., $\Delta x_{r_i} \ll L$. If the radius of the outer sphere is much less than the length of the pendulum and the motion of the resonant mass is constrained within the outer sphere, then this condition is, in general, met. However, since in this study we are also only investigating systems with the condition $\omega_r \gg \omega_g$ (i.e., internal resonator frequency $\omega_r=\sqrt{k_r/m_r}$ much greater than the pendulum frequency $\omega_g=\sqrt{k_g/m}$), the condition $\Delta x_{r_i} \ll L$ is further strengthened. This is confirmed by the simulation results presented in Fig.~5 of the main text. In such a case, the displacement and resulting velocity and acceleration vectors reduce to
$$
\begin{aligned}
& \left[\begin{array}{c}x_i \\ y_i \end{array}\right]=\left[\begin{array}{c}
L \theta_i \\
0
\end{array}\right], \\
& \left[\begin{array}{c}x_{r_i} \\ y_{r_i} \end{array}\right]
=\left[\begin{array}{c}
x_i + \Delta x_{r_i} \\
0
\end{array}\right], \\
& \left[\begin{array}{c}\dot{x_i} \\ \dot{y_i} \end{array}\right]=\left[\begin{array}{c}
L \dot{\theta_i} \\
0
\end{array}\right], \\
& \left[\begin{array}{c}\dot{x}_{r_i} \\ \dot{y}_{r_i} \end{array}\right]
=\left[\begin{array}{c}
\dot{x}_i + \Delta \dot{x}_{r_i} \\
0
\end{array}\right], \\
& \left[\begin{array}{c}\ddot{x_i} \\ \ddot{y_i} \end{array}\right]=\left[\begin{array}{c}
L \ddot{\theta_i} \\
0
\end{array}\right], \\
& \left[\begin{array}{c}\ddot{x}_{r_i} \\ \ddot{y}_{r_i} \end{array}\right]
=\left[\begin{array}{c}
\ddot{x_i} + \Delta \ddot{x}_{r_i} \\
0
\end{array}\right].
\end{aligned}
$$
Substituting these expressions into the equation of motion Eq.~(\ref{suppeq_mnew}) and dividing through by $L$ leads to
\begin{equation}
m \ddot{x_i}+m_r \frac{\Delta x_{r_i}}{L^2}\left(\ddot{x}_i \Delta x_{r_i}+2 \dot{x}_i \Delta \dot{x}_{r_i}+\dot{x}_i^{2}\right)+m_r g \frac{\Delta x_{r_i}}{L}+\frac{mg}{L}x_i-k_{r}\left(x_{r_i}-x_i\right)=0. \nonumber
\end{equation}
Substituting into Eq.~(\ref{suppeq_mr}) for the inner resonant sphere leads to
\begin{equation}
m_r\ddot{x}_{r_i}-\frac{\Delta x_{r_i}}{L^2} \dot{x}_i^{2}+\frac{m_{r}g}{L} x_i+k_{r}\left(x_{r_i}-x_i\right)=0. \nonumber
\end{equation} 
For $\Delta x_{r_i} \ll L$, these two equations of motion reduce to the following one-dimension approximations:
\begin{eqnarray}
& m \ddot{x_i}+\frac{mg}{L}x_i-k_{r}\left(x_{r_i}-x_i\right)=0, \nonumber \\
& m_r\ddot{x}_{r_i}+\frac{m_{r}g}{L} x_i+k_{r}\left(x_{r_i}-x_i\right)=0. \nonumber
\end{eqnarray}
Defining an effective pendulum linear spring constant $k_g = {mg}/{L}$ leads to the following equations of motion:
\begin{eqnarray}
\Aboxed{& m \ddot{x_i}+k_g x_i-k_{r}\left(x_{r_i}-x_i\right)=0,} \\
\Aboxed{& m_r\ddot{x}_{r_i}+\frac{m_{r}}{m} k_g x_i+k_{r}\left(x_{r_i}-x_i\right)=0.}
\end{eqnarray}
\end{section}

\begin{section}{Derivation of the instantaneous effective coefficients of restitution $\mathit{CR_e}$ and motion $\mathit{CM_e}$}
The separation velocities for the two-spherical shell mass-in-mass resonator system in the time domain can be expressed in terms of an instantaneous effective coefficient of restitution $\mathit{CR_e}$ and an effective coefficient of motion $\mathit{CM_e}$:
$$
\begin{aligned}
&\dot{x}_{1,t_s} = \frac{1}{2}\left(\mathit{CM}_e-\mathit{CR}_e\right)\dot{x}_{1,t_c}, \\
&\dot{x}_{2,t_s} = \frac{1}{2}\left(\mathit{CM}_e+\mathit{CR}_e\right)\dot{x}_{1,t_c}.
\end{aligned}
$$
This formulation is equivalent to the case for identical conventional spherical shells when $\mathit{CM_e}=1$:
$$
\begin{aligned}
&\dot{x}_{1,t_s} = \frac{1}{2}\left(1-\mathit{CR}_e\right)\dot{x}_{1,t_c}, \\
&\dot{x}_{2,t_s} = \frac{1}{2}\left(1+\mathit{CR}_e\right)\dot{x}_{1,t_c}.
\end{aligned}
$$
The effective coefficient of restitution $\mathit{CR_e}$ can be defined in an equivalent manner to the conventional case, based on the ratio of the separation velocity to the incident collision velocity, noting that the second mass-in-mass resonator is stationary when the collision starts (i.e., corresponding to the standard Newton's-cradle initial condition for a first collision):
$$
\mathit{CR_e}=\frac{\dot{x}_{2, t_s}-\dot{x}_{1, t_s}}{\dot{x}_{1, t_c}}.
$$
Accordingly, the effective coefficient of motion $\mathit{CM_e}$ is given by
$$
\mathit{CM_e}=\frac{\dot{x}_{1, t_s}+\dot{x}_{2, t_s}}{\dot{x}_{1, t_c}}.
$$
To derive the effective collision parameters, we begin with the equations of motion in the time domain during the time interval when the mechanically identical mass-in-mass resonators are in collision. The compression force of the shells is modeled using a linear approximation:
$$
\begin{aligned}
\text{Spherical shell 1:}\\
& m \ddot{x}_{1}=k_c\left(x_{2}-x_{1}\right)+k_{r}\left(x_{r_1}-x_{1}\right)-k_g x_{1}, \\
& m_r\ddot{x}_{r_1}=k_r\left(x_{1}-x_{r_1}\right) - \frac{m_{r}}{m} k_g x_1.\\
\text{Spherical shell 2:} \\
& m \ddot{x}_{2}=k_c\left(x_{1}-x_{2}\right)+k_{r}\left(x_{r_2}-x_{2}\right)-k_g x_{2}, \\
& m_r\ddot{x}_{r_2}=k_r\left(x_{2}-x_{r_2}\right) - \frac{m_{r}}{m} k_g x_2. \\
\end{aligned}
$$
We start by simplifying the equations of motion during the collision by assuming that the resonator stiffness is of a similar order of magnitude to the compression stiffness in the conventional case, i.e., $k_r \sim k_c$ and $k_{c} \gg k_{g}$:
$$
\begin{aligned}
\text{Spherical shell 1:}\\
& m \ddot{x}_{1}=k_c\left(x_{2}-x_{1}\right)+k_{r}\left(x_{r_1}-x_{1}\right), \\
& m_r\dot{x}_{r_1}=k_r\left(x_{1}-x_{r_1}\right). \\
\text{Spherical shell 2:} \\
& m \ddot{x}_{2}=k_c\left(x_{1}-x_{2}\right)+k_{r}\left(x_{r_2}-x_{2}\right), \\
& m_r\ddot{x}_{r_2}=k_r\left(x_{2}-x_{r_2}\right). \\
\end{aligned}
$$
Consider the first mass-in-mass resonator. We convert to the Laplace domain (retaining the initial conditions at the point of impact) for ease of algebraic manipulation:
$$
\begin{aligned}
& m\left(s^{2} X_{1}-s x_{1,t _c}-\dot{x}_{1,t_c}\right)=k_c\left(X_{2}-X_{1}\right)+k_{r}\left(X_{r_1}-X_{1}\right), \\
& m_r\left(s^{2} X_{r_1}-s x_{r_1,t_c}-\dot{x}_{r,t_c}\right)=k_{r}\left(X_{1}-X_{r_1}\right). \\
\end{aligned}
$$
Lumping the initial conditions for the outer mass in the quantity $g_{1}=s x_{1,t _c}+\dot{x}_{1,t_c}$ and resonator mass in the quantity $g_{r_1}=s x_{r_1,t_c}+\dot{x}_{r,t_c}$ for compactness,
$$
\begin{aligned}
& m\left(s^{2} X_{1}-g_{1}\right)=k_c\left(X_{2}-X_{1}\right)+k_{r}\left(X_{r_1}-X_{1}\right), \\
& m_{r}\left(s^{2} X_{r_{1}}-g_{r_{1}}\right)=k_{r}\left(X_{1}-X_{r_1}\right).\\
\end{aligned}
$$
Rearranging leads to Eq.~(\ref{suppeq}) and Eq.~(\ref{suppeq1}):
\begin{equation}
\left(m s^{2}+k_{r}\right) X_{1}=k_{r} X_{r_1}+k_c\left(X_{2}-X_{1}\right)+m g_1,   
\label{suppeq}
\end{equation}
\begin{equation}
\left(m_r s^{2}+k_{r}\right) X_{r_{1}}=k_{r} X_{1}+m_r g_{r_{1}}.   \\
\label{suppeq1}
\end{equation}\\
\text{Substituting for $X_{r_1}$ from Eq.~(\ref{suppeq1}) into Eq.~(\ref{suppeq}) and rearranging leads to,}
$$
\begin{aligned}
\left(m s^{2}+k_{r}\right) X_{1} &= k_{r} \frac{\left(k_{r} X_{1}+m_{r} g_{r}\right)}{m_{r} s^{2}+k_{r}} + k_c\left(X_{2}-X_{1}\right) + m g_{1},\\
\\
\left(m s^{2}+k_r - \frac{k_{r}^{2}}{m_r s^{2}+k_r}\right) X_{1} &= \frac{k_r m_r g_{r}}{m_r s^{2}+k_{r}} + k_c\left(X_{2}-X_{1}\right) + m g_1,\\
\left(m + \frac{k_r m_r}{m_r s^{2}+k_r}\right) s^{2} X_{1} &= k_c\left(X_{2}-X_{1}\right) + \frac{k_r m_r g_{r}}{m_r s^{2}+k_{r}} + m g_1.
\end{aligned}
$$
Defining the effective mass in the same manner as in previous work on mass-in-mass resonators in the field of metamaterials,
$$
\begin{aligned}
& m_e = m+\frac{k_r m_r}{m_r s^{2}+k_r},\\
\end{aligned}
$$
leads to
$$
\begin{aligned}
& m_e s^{2} X_{1}=k_c\left(X_{2}-X_{1}\right)+\frac{k_{r} m_{r} g_{r_1}}{m_r s^{2}+k_r}+m g_{1}.
\end{aligned}
$$
Through a similar approach, the equation of motion of the second mass-in-mass resonator in the Laplace domain can be expressed as follows:
$$
m_e s^{2} X_{2}=k_c\left(X_{1}-X_{2}\right)+\frac{k_r m_r g_{r_2}}{m_rs^{2}+k_r}+m g_{2}.
$$
For the initial conditions appropriate to a conventional standard Newton's cradle, $g_{2}=g_{r_{2}}=0$, the equation of motion becomes
$$
m_e s^{2} X_{2}=k_c\left(X_{1}-X_{2}\right).
$$
The equations of motion in the Laplace domain can therefore be written as follows:
\begin{equation}
m_es^{2} X_{1}=k_c\left(X_{2}-X_{1}\right)+g_{\rm{all}},
\label{suppeq2}
\end{equation}
\begin{equation}
m_es^{2} X_{2}=k_c\left(X_{1}-X_{2}\right), 
\label{suppeq3}
\end{equation}
\text{where}\\
\begin{equation}
\nonumber
 g_{\rm{all}}=\frac{k_{r} m_{r} g_{r_1}}{m_r s^{2}+k_r}+m g_{1}.
\end{equation}\
\\
Equations~(\ref{suppeq2}) and (\ref{suppeq3}) lead to the following expressions:
$$
\begin{aligned}
m_es^{2}\left(X_{2}-X_{1}\right) & =k_c\left(X_{1}-X_{2}-X_{2}+X_{1}\right)-g_{\rm{all}}, \\
& =-2 k_c\left(X_{2}-X_{1}\right)-g_{\rm{all}},\\
\left(m_es^{2}+2 k_c\right)\left(X_{2}-X_{1}\right) & =-g_{\rm{all}}, \\
X_{2}-X_{1}&=\frac{-g_{\rm{all}}}{m_{e} s^{2}+2 k_c}.
\end{aligned}
$$
Using the earlier definition of the effective coefficient of restitution, and substituting for the spherical-shell separation velocity leads to the following equation:
$$
\mathit{CR_e}=\frac{\dot{x}_{2, t_s}-\dot{x}_{1, t_s}}{\dot{x}_{1, t_c}}=\frac{\left.\mathcal{L}^{-1} s\left(X_{2}-X_{1}\right)\right|_{t=t_s}}{\dot{x}_{1, t_c}},\\
$$
where $\mathcal{L}^{-1} $ is the inverse Laplace transform. We shall later make use of the following notation: $\mathit{CR_e}$ is defined as $\mathit{CR_e}(t)$ at $t=t_s$, i.e., $\mathit{CR_e}(t_s)$.
Writing the effective mass as
$$
m_{e}=m+m_{e, s},
$$
the lumped initial-condition term can be written in the following form:
$$
\begin{aligned}
g_{\rm{all}}&=m_{e ,s} g_{r_1}+m g_1,\\
& =m_{e, s} g_{r_1}+m \dot{x}_{1, t_c}.
\end{aligned}
$$
The conditions for spherical-shell separation in the Laplace domain can then be expressed in the form
$$
\begin{aligned}
&X_{2}-X_{1}=\frac{-(m_e-m) g_{r_1}-m \dot{x}_{1, t_c}}{m_e s^{2}+2 k_c}, \\
& =\frac{-(m_e-m)\left(s x_{r_1, t_c}+\dot{x}_{r_1, t_c}\right)-m \dot{x}_{1, t_c}}{m_e s^{2}+2 k_c}. \\
\end{aligned}
$$
Defining the effective mass in terms of the resonator frequency, one obtains
$$
\hat{m}_{e}=\frac{m_e}{m}=1+\frac{k_r / m}{s^{2}+\omega_r^{2}}=1+\frac{m_r}{m} \frac{\omega_r^{2}}{s^{2}+\omega_r^{2}}=1+\hat{m}_{e,s},\\
$$
Together with $\omega_c = \sqrt{2k_c/m}$, leads to
$$
\begin{aligned}
X_{2}-X_{1}&=\frac{-m\left(\hat{m}_{e}-1\right) g_r-m \dot{x}_{1, t_c}}{m \hat{m}_{e}s^{2}+m \omega_{c}^{2}},\\ 
& =\frac{-\left(\hat{m}_{e}-1\right) g_{r}-\dot{x}_{1, t c}}{\hat{m}_e s^{2}+\omega_c^{2}},\\
& =\frac{-\hat{m}_{e,s} g_{r}-\dot{x}_{1, t c}}{\hat{m}_e s^{2}+\omega_c^{2}},\\
& =-\frac{g_{\rm{all}}}{m}\frac{1}{\hat{m}_e s^{2}+\omega_c^{2}}.\\
\end{aligned}
$$
Substitution into the effective coefficient of restitution gives
$$
\begin{aligned}
\mathit{CR_e} & =-\left.\mathcal{L}^{-1} \left(\frac{g_{\rm{all}}}{m\dot{x}_{1, t_c}}\frac{s}{\hat{m}_e s^{2}+\omega_c^{2}}\right)\right|_{t=t_s},\\
& =-\left.\mathcal{L}^{-1} s\frac{\hat{m}_{e,s}\frac{g_r}{\dot{x}_{1, t_c}}+1}{\hat{m}_e s^{2}+\omega_c^{2}}\right|_{t=t_s}.\\
\\
\Aboxed{\mathit{CR_e} & =-\left.\mathcal{L}^{-1} \left(\frac{\hat{m}_{e,s}s}{\hat{m}_e s+\omega_c^{2}}\left(\frac{x_{r_1,t_c}}{\dot{x}_{1, t_c}}s+\frac{\dot{x}_{r,t_c}}{\dot{x}_{1, t_c}}\right) +\frac{s}{\hat{m}_e s^{2}+\omega_c^{2}}\right)\right|_{t=t_s}.}
\end{aligned}
$$
In a similar manner one can define the effective coefficient of motion $\mathit{CM_e}$ in the time domain. For later use, $\mathit{CM_e}$ is defined as $\mathit{CM_e}(t)$ at $t=t_s$, i.e., $\mathit{CM_e}(t_s)$.

From Eqs.~(\ref{suppeq2}) and (\ref{suppeq3}),
$$
m_e{s}^{2}\left(X_{1}+X_{2}\right)=g_{\rm{all}}.
$$
Substituting into the expression for $\mathit{CM_e}$, one obtains
$$
\begin{aligned}
\mathit{CM_e}&=\frac{\dot{x}_{1, t_s}+\dot{x}_{2, t_s}}{\dot{x}_{1, t_c}},\\
& =\left.\frac{\mathcal{L}^{-1} s\left(X_{1}+X_{2}\right)}{\dot{x}_{1, t_c}}\right|_{t=t_s},\\
& =\left.\mathcal{L}^{-1}\left(\frac{g_{\rm{all}}}{m_{e} s \dot{x}_{1, t_c}}\right)\right|_{t=t_{s}},\\
& =\left.\mathcal{L}^{-1}\left(\frac{g_{\rm{all}}}{m\dot{x}_{1, t_c}}\frac{1}{\hat{m}_{e} s}\right)\right|_{t=t_{s}},\\
& =\left.\mathcal{L}^{-1}\left(\frac{\hat{m}_{e, s} g_{r_1}+\dot{x}_{1, t_c}}{\hat{m}_{e} s \dot{x}_{1, t_c}}\right)\right|_{t=t_{s}},\\
& =\left.\mathcal{L}^{-1}\left(\frac{\hat{m}_{e, s} g_{r_1}}{\hat{m}_{e} s \dot{x}_{1, t_c}}+\frac{1}{\hat{m}_{e} s}\right)\right|_{t=t_{s}}.\\
\\
\Aboxed{\mathit{CM_e} & =\left.\mathcal{L}^{-1}\left(\frac{\hat{m}_{e, s}}{\hat{m}_{e} s}\left(s \frac{x_{r_1,t_c}}{ \dot{x}_{1, t_c}}+\frac{\dot{x}_{r,t_c}}{ \dot{x}_{1, t_c}}\right)+\frac{1}{\hat{m}_{e} s}\right)\right|_{t=t_{s}}.}
\end{aligned}
$$
\end{section}

\begin{section}{Derivation of the post-collision velocity in the external potential}

The equations of motion of spherical shell $i$ during free oscillation are given by

$$
\begin{aligned}
& m \ddot{x}_{i}=-k_g x_{i}+k_r\left(x_{r_{i}}-x_{i}\right),\\
& m_r \ddot{x}_{r_i}=-\frac{m_{r}}{m} k_g x_i+k_r\left(x_{i}-x_{r_i}\right).
\end{aligned}
$$
Converting to Laplace domain, with the initial conditions taken at $t_s$, the moment at which the shells separate after the collision,
$$
\begin{aligned}
& m\left(s^{2} X_{i}-s x_{i,t_s}-\dot{x}_{i,t_s}\right)=-k_{g} X_{i}+k_{r}\left(X_{r_i}-X_{i}\right),\\
& m_{r}\left(s^{2} X_{r_i}-s x_{r_i,t_s}-\dot{x}_{r_i,t_s}\right)=-\frac{m_{r}}{m} k_g X_i+k_{r}\left(X_{i}-X_{r_i}\right),\\
& \\
& \left(s^{2} X_{i}-g_{i,t_s}\right)=-\omega_{g}^{2} X_{i}+\frac{m_{r}}{m} \omega_{r}^{2}\left(X_{r_i}-X_{i}\right),\\
& \left(s^{2} X_{r_i}-g_{r_i,t_s}\right)=-\omega_{g}^{2} X_{i}\omega_{r}^{2}\left(X_{i}-X_{r_i}\right),\\
\end{aligned}\\
$$
\begin{equation}
\left(s^{2}+\frac{m_r}{m} \omega_{r}^{2}+\omega_{g}^{2}\right) X_{i}=g_{i,t_s}+\frac{m_r}{m} \omega_{r}^{2} X_{r_i},\\
\label{suppeq4}
\end{equation}\\
\begin{equation}
\left(s^{2}+\omega_{r}^{2}+\omega_{g}^{2}\right) X_{r_{i}}=g_{r_i,t_s}+\omega_{r}^{2} X_{i}.
\label{suppeq5}
\end{equation}
From Eq.~(\ref{suppeq5}),
$$
X_{r_i}=\frac{g_{r_i,t_s}+\omega_{r}^{2} X_{i}}{s^{2}+\omega_{r}^{2}+\omega_{g}^{2}}.
$$
Substituting in Eq.~(\ref{suppeq4}) yields
$$
\left(s^{2}+\frac{m_{r}}{m} \omega_{r}^{2}+\omega_{g}^{2}\right) X_{i}=g_{i, t_s}+\frac{m_{r}}{m} \omega_{r}^{2}\frac{g_{r_i, t_{s}}+\omega_{r}^{2}X_{i}}{s^2+\omega_{r}^{2}+\omega_{g}^{2}}.
$$
Rearranging to solve for $X_i$, one obtains
$$
\begin{aligned}
&\left(\left(s^{2}+\frac{m_r}{m} \omega_{r}^{2}+\omega_{g}^{2}\right)\left(s^{2}+\omega_{r}^{2}+\omega_{g}^{2}\right)-\frac{m_{r}}{m} \omega_{r}^{4}\right) X_{i} =g_{i, t_s}\left(s^{2}+\omega_{r}^{2}+\omega_{g}^{2}\right)+g_{r_i, t_s} \frac{m_{r}}{m} \omega_{r}^{2},
\end{aligned}
$$
$$
\begin{aligned}
X_{i}&=\frac{\left(s^{2}+\omega_{r}^{2}+\omega_{g}^{2}\right) g_{i, t_s}+\frac{m_{r}}{m} \omega_{r}^{2} g_{r_i,t_s}}{\left(s^{2}+\frac{m_{r}}{m} \omega_{r}^{2}+\omega_{g}^{2}\right)\left(s^{2}+\omega_{r}^{2}+\omega_{g}^{2}\right)-\frac{m_{r}}{m} \omega_{r}^{4}},\\
& =\frac{\left(s^{2}+\omega_{r}^{2}+\omega_{g}^{2}\right) g_{i, t_s}+\frac{m_r}{m}\omega_{r}^{2} g_{r_{i}, t_s}}{s^{4}+\left(\frac{m+m_r}{m} \omega_{r}^{2}+ 2\omega_{g}^{2}\right)s^{2}+\frac{m_r}{m}\omega_{r}^{4}+\omega_g^2\omega_r^2+\omega_{g}^{4}+\frac{m_r}{m}\omega_{g}^{2}\omega_r^2-\frac{m_r}{m}\omega_{r}^{4}},\\
& =\frac{\left(s^{2}+\omega_{r}^{2}+\omega_{g}^{2}\right) g_{i, t_s}+\frac{m_r}{m}\omega_{r}^{2} g_{r_i, t_s}}{s^{4}+\left(\frac{m+m_r}{m} \omega_{r}^{2}+2\omega_{g}^{2}\right) s^{2}+\frac{m+m_r}{m}\omega_{g}^{2}\omega_{r}^{2}+\omega_{g}^{2}}, \\
& =\frac{\left(s^{2}+\omega_{r}^{2}+\omega_{g}^{2}\right) g_{i, t_s}+\frac{m_r}{m}\omega_{r}^{2} g_{r_i, t_s}}{\left(s^{2}+\omega_g^2\right)\left(s^2+\frac{m+m_r}{m} \omega_{r}^{2}+\omega_{g}^{2}\right)}, \\
\end{aligned}
$$
Using partial factor decomposition, we can write $X_i$ in terms of two fundamental angular frequencies  $\omega_{g}$ and $\omega_{r^\prime}$:
$$
\begin{aligned}
X_{i}&=\frac{A}{s^{2}+\omega_{g}^{2}}+\frac{B}{s^{2}+\omega_{r^\prime}^{2}}=\frac{s^{2}\left(A+B\right)+A \omega_{r^\prime}^{2}+B\omega_{g}^{2}}{\left(s^{2}+\omega_{g}^{2}\right) \left(s^{2}+\omega_{r^\prime}^{2}\right)}, \\
\text{with,}\\
\omega_{r^\prime}^{2}&=\frac{m+m_r}{m} \omega_{r}^{2}+\omega_{g}^{2}.
\end{aligned}
$$
Comparing coefficients to determine the numerators,
$$
\begin{aligned}
&s^{2}: &A+B=&g_{i,t_s} \\
&s^{0}: &A \omega_{r^\prime}^{2}+B \omega_{g^\prime}^{2}=&\left(\omega_{r}^{2}+\omega_{g}^{2}\right) g_{i, t_s}+\frac{m_r}{m}\omega_{r}^{2} g_{r_i, t_s}.
\end{aligned}
$$
From $s^{2}$ we get $B=g_{i,t_s}-A$, and substituting into $s^0$ yields
$$
\begin{aligned}
A\omega_{r^\prime}^{2}+\left(g_{i,t_s}-A\right) \omega_{g}^{2}&=\left(\omega_{r}^{2}+\omega_{g}^{2}\right) g_{i, t_s}+\frac{m_r}{m}\omega_{r}^{2} g_{r_i, t_s},\\
A&=\frac{\left(\omega_{r}^{2}+\omega_{g}^{2}\right) g_{i, t_s}+\frac{m_r}{m}\omega_{r}^{2} g_{r_i,t_{s}}-\omega_{g}^{2} g_{i, t_s}}{\omega_{r^\prime}^2-\omega_{g}^2},\\
& =\frac{\omega_{r}^{2} g_{i, t_s}+\frac{m_r}{m}\omega_{r}^{2} g_{r_i, t_s}}{\frac{m+m_r}{m} \omega_{r}^{2}},\\
& =\frac{m g_{i, t_s}+m_r g_{r_i, t_s}}{\left(m+m_r\right)},\\
& =\frac{m\dot{x}_{i, t_s}+m_r \dot{x}_{r_i, t_s}}{m+m_{r}}+\frac{m x_{i, t_s}+m_{r}x_{r_i, t_s}}{m+m_{r}}s,\\
\end{aligned}
$$
and
$$
\begin{aligned}
B&=g_{i,t_{s}}-A,\\
& =\frac{m_r\left(g_{i, t_s}+g_{r_i, t_s}\right)}{\left(m+m_r\right)}\\
& =\frac{m_r}{m+m_{r}}\left(\dot{x}_{i, t_s}-\dot{x}_{r_i, t_s}\right)+\frac{m_r}{m+m_r}\left(x_{i, t_s}-x_{r_i, t_s}\right)s.
\end{aligned}
$$
Since $\omega_r \gg \omega_g$ for the system studied,
$$
\begin{aligned}
\omega_{r^\prime}^{2}&= \frac{m+m_r}{m}\omega_{r}^{2}.\\\\
\end{aligned}
$$
This leads to the time-domain velocity
$$
\begin{aligned}
\dot{x}_{i}\left(t\right) =&\mathcal{L}^{-1}\left(s X_{i}\right),\\
 =&\left(\frac{m \dot{x}_{i,t_s}+m_{r} \dot{x}_{r_i, t_s}}{m+m_r}\right) \cos \omega_g t-\left(\frac{m x_{i, t_s}+m_r x_{i, t_s}}{m+m_r}\right)  \omega_g \sin \omega_g t,\\
& +\frac{m_r}{m+m_r}\left(\dot{x}_{i, t_s}-\dot{x}_{r_i, t_s}\right) \cos \sqrt{\frac{m+m_r}{m}} \omega_r t-\frac{m_r}{m+m_r}\left(x_{i, t_s}-x_{r_i, t_s}\right) \sqrt{\frac{m+m_r}{m}} \omega_r \sin \sqrt{\frac{m+m_r}{m}} \omega_r t.
\end{aligned}
$$
\end{section}

\begin{section}{Derivation of the overall effective coefficient of restitution $\mathit{CR_a}$ and effective coefficient of motion $\mathit{CM_a}$ for the average post-collision motion}

The average post-collision velocities in the external potential owing to the sum of the pendulum and internal resonator motion are defined as
$$
\begin{aligned}
\dot{x}_{1,a} =& \frac{1}{2}\left(\mathit{CM_a}-\it{CR}_a\right)\dot{x}_{1,t_c},\\
\dot{x}_{2,a} =& \frac{1}{2}\left(\mathit{CM_a}+\it{CR}_a\right)\dot{x}_{1,t_c},\\ \\
\text{with} \\
\\
\it{CR}_{a} =& \frac{1}{m+m_r}\frac{m\left(\dot{x}_{2,t_s}-\dot{x}_{1,t_s}\right)+m_r\left(\dot{x}_{r,2,t_s}-\dot{x}_{r,1,t_s}\right)}{\dot{x}_{1,t_c}},\\
=& \frac{1}{m+m_r}\left(m\it{CR}_e + m_r\it{CR}_{r}\right),\\
\mathit{CM_a} =& \frac{1}{m+m_r}\frac{m\left(\dot{x}_{2,t_s}+\dot{x}_{1,t_s}\right)+m_r\left(\dot{x}_{r,2,t_s}+\dot{x}_{r,1,t_s}\right)}{\dot{x}_{1,t_c}},\\
=& \frac{1}{m+m_r}\left(m\mathit{CM_e} + m_r\mathit{CM_r}\right) \label{eq:m_effective_a}.\\
\end{aligned}
$$
Following a similar approach to that previously, the equations of motion of the resonator masses during collision in the Laplace domain can be written as follows:
\begin{equation}
m_{r}\left(s^{2} X_{r_{1}}-g_{r_{1}}\right)=k_{r}\left(X_{1}-X_{r_{1}}\right),\\
\label{suppeq6}
\end{equation}
\begin{equation}
m_{r}s^{2} X_{r_{2}}=k_{r}\left(X_{2}-X_{r_{2}}\right).
\label{suppeq7}
\end{equation}
\text{From Eqs.~(\ref{suppeq6}) and (\ref{suppeq7}),}
$$
\begin{aligned}
m_rs^2\left(X_{r_2}-X_{r_1}\right) & =-k_{r}\left(X_{r_{2}}-X_{r_{1}}\right)+k_{r}\left(X_{2}-X_{1}\right)-m_r g_{r},\\
X_{r_{2}}-X_{r_{1}} & =\frac{k_{r}\left(X_{2}-X_{1}\right)-m_r g_{r}}{m_{r}s^{2}+k_{r}},\\
& =m_{e, s}\left(\frac{X_{2}-X_{1}}{m_{r}}-\frac{g_{r}}{k_{r}}\right),\\
& =\hat{m}_{e,s}\left(\frac{m}{m_{r}}\left(X_{2}-X_{1}\right)- \frac{m}{k_{r}} g_{r}\right),\\
& =\frac{m}{m_r}\hat{m}_{e,s}\left(\left(X_{2}-X_{1}\right)- \omega_r^2 g_{r}\right),\\
& =\frac{m}{m_r}\left(\hat{m}_{e}-1\right)\left(\left(X_{2}-X_{1}\right)- \omega_r^2 g_{r}\right).
\end{aligned}
$$
The resonator effective coefficient of restitution is defined as
$$
\mathit{CR_r}=\frac{\dot{x}_{r_2, t_s}-\dot{x}_{r_1, t_s}}{\dot{x}_{1, t_c}}=\frac{\left.\mathcal{L}^{-1} s\left(X_{r_2}-X_{r_1}\right)\right|_{t=t_s}}{\dot{x}_{1, t_c}}.\\
$$
Substituting for $X_{r_2}-X_{r_1}$ leads to
$$
\begin{aligned}
\mathit{CR_{r}}&=\frac{\frac{m}{m_r} \left.\left(\mathcal{L}^{-1} \hat{m}_e s\left(X_{2}-X_{1}\right)-\mathcal{L}^{-1} s\left(X_{2}-X_{1}\right)-\omega_r^2 \mathcal{L}^{-1}g_r s\left(\hat{m}_{e}-1\right)\right)\right|_{t=t_s}}{\dot{x}_{1, t_{c}}},\\
& =\frac{m}{m_r}\left(\left.\left(\hat{m}_{e}(t) * \mathit{CR_{e}}(t)\right)\right|_{t=t_{s}}-\mathit{CR_{e}}-\left.\omega_r^2\left(\frac{d^{2} \hat{m}_{e}(t)}{d t^{2}}\frac{x_{r_1, t_{c}}}{\dot{x}_{1, t_c}}+\frac{d \hat{m}_{e}(t)}{d t}\frac{\dot{x}_{r_1, t_{c}}}{\dot{x}_{1, t_c}}\right)\right|_{t=t_{s}}\right),\\
& =\frac{m}{m_r}\left(\left.\left(\hat{m}_{e,s}(t) * \mathit{CR_{e}}(t)\right)\right|_{t=t_{s}}-\left.\omega_r^2\left(\frac{d^{2} \hat{m}_{e,s}(t)}{d t^{2}}\frac{x_{r_1, t_{c}}}{\dot{x}_{1, t_c}}+\frac{d \hat{m}_{e,s}(t)}{d t}\frac{\dot{x}_{r_1, t_{c}}}{\dot{x}_{1, t_c}})\right)\right|_{t=t_{s}}\right).
\end{aligned}
$$
From Eqs.~(\ref{suppeq6}) and (\ref{suppeq7}),
$$
\begin{aligned}
m_{r} s^{2}\left(X_{r_1}+X_{r_{2}}\right)&=k_{r}\left(X_{1}+X_{2}-X_{r_{1}}-X_{r_{2}}\right)+m_{r} g_{r_1},\\
X_{r_{1}}+X_{r_{2}}&=\frac{k_{r}\left(X_{1}+X_{2}\right)+m_{r} g_{r_1}}{m_{r} s^{2}+k_{r}},\\
& =m_{e,s}\left(\frac{X_{1}+X_{2}}{m_{r}}+\frac{g_{r_{1}}}{k_r}\right),\\
& =\frac{m}{m_{r}}\hat{m}_{e,s}\left(\left(X_{1}+X_{2}\right)+\frac{g_{r_{1}}}{\omega_r^2}\right).\\
\end{aligned}
$$
The resonator effective coefficient of motion is defined as
$$
\begin{aligned}
\mathit{CM_r}&=\frac{\left.\mathcal{L}^{-1} s\left(X_{r_{1}}+X_{r_{2}}\right)\right|_{t=t_s}}{\dot{x}_{1, t_c}},\\
&=\frac{m}{m_r}\left(\left.\left(\hat{m}_{e, s}(t) * \mathit{CM_e}(t)\right)\right|_{t=t_{s}}+\frac{1}{\omega_r^2}\left.\left(\frac{d^{2} \hat{m}_{e, s}(t)}{d t^{2}} \frac{x_{r_1, t_{c}}}{\dot{x}_{1, t_c}}+\frac{d {\hat{m}_{e, s}}(t)}{d t} \frac{\dot{x}_{r_1, t_c}}{\dot{x}_{1, t_c}}\right)\right|_{t=t_{s}}\right).
\end{aligned}
$$
\newline
\newline
From the definition of $\it{CR}_a$,
$$
\begin{aligned}
\it{CR}_{a} =& \frac{1}{m+m_r}\frac{m\left(\dot{x}_{2,t_s}-\dot{x}_{1,t_s}\right)+m_r\left(\dot{x}_{r,2,t_s}-\dot{x}_{r,1,t_s}\right)}{\dot{x}_{1,t_c}},\\
=& \frac{1}{m+m_r}\left(m\it{CR}_e + m_r\it{CR}_{r}\right) \label{eq:CR_effective_a}.\\
\end{aligned}
$$
Substituting for $\it{CR}_e$ and $\it{CR}_r$ leads to
$$
\begin{aligned}
\it{CR}_{a} = & \frac{m}{m+m_r}\left(\left.\left(\hat{m}_{e}(t) * \mathit{CM_e}(t)\right)\right|_{t=t_{s}}-\left.\omega_r^2\left(\frac{d^{2} \hat{m}_{e}(t)}{d t^{2}}\frac{x_{r_1, t_{c}}}{\dot{x}_{1, t_c}}+\frac{d \hat{m}_{e}(t)}{d t}\frac{\dot{x}_{r_1, t_{c}}}{\dot{x}_{1, t_c}}\right)\right|_{t=t_{s}}\right).\\
\Aboxed{\it{CR}_{a} = & \frac{m}{m+m_r}\left(CR_{e} + \left.\left(\hat{m}_{e,s}(t) * CR_{e}(t)\right)\right|_{t=t_{s}}-\left.\omega_r^2\left(\frac{d^{2} \hat{m}_{e,s}(t)}{d t^{2}}\frac{x_{r_1, t_{c}}}{\dot{x}_{1, t_c}}+\frac{d \hat{m}_{e,s}(t)}{d t}\frac{\dot{x}_{r_1, t_{c}}}{\dot{x}_{1, t_c}}\right)\right|_{t=t_{s}}\right).}\\
\end{aligned}
$$
\newline
\newline
From the definition of $\it{CM}_a$,
$$
\begin{aligned}
\it{CM}_{a} =& \frac{1}{m+m_r}\frac{m\left(\dot{x}_{2,t_s}+\dot{x}_{1,t_s}\right)+m_r\left(\dot{x}_{r,2,t_s}+\dot{x}_{r,1,t_s}\right)}{\dot{x}_{1,t_c}},\\
=& \frac{1}{m+m_r}\left(m\it{CM}_e + m_r\it{CM}_{r}\right) \label{eq:M_effective_a}.\\
\end{aligned}
$$
Substituting for $\it{CM}_e$ and $\it{CM}_r$ leads to
$$
\begin{aligned}
\it{CM}_{a} =& \frac{m}{m+m_r}\left(\it{CM}_e + \left.\left(\hat{m}_{e, s}(t) * \mathit{CM_e}(t)\right)\right|_{t=t_{s}}+\frac{1}{\omega_r^2}\left.\left(\frac{d^{2} \hat{m}_{e, s}(t)}{d t^{2}} \frac{x_{r_1, t_{c}}}{\dot{x}_{1, t_c}}+\frac{d {\hat{m}_{e, s}}(t)}{d t} \frac{\dot{x}_{r_1, t_c}}{\dot{x}_{1, t_c}}\right)\right|_{t=t_{s}}\right),\\
\it{CM}_{a} =& \frac{m}{m+m_r}\left(\left.\left(\hat{m}_{e}(t) * \mathit{CM_e}(t)\right)\right|_{t=t_{s}}+\frac{1}{\omega_r^2}\left.\left(\frac{d^{2} \hat{m}_{e, s}(t)}{d t^{2}} \frac{x_{r_1, t_{c}}}{\dot{x}_{1, t_c}}+\frac{d {\hat{m}_{e, s}}(t)}{d t} \frac{\dot{x}_{r_1, t_c}}{\dot{x}_{1, t_c}}\right)\right|_{t=t_{s}}\right),\\
\it{CM}_{a} =& \frac{m}{m+m_r}\left(\left.\mathcal{L}^{-1} \hat{m}_{e}\left(\frac{\hat{m}_{e, s}}{\hat{m}_{e} s}\left(s \frac{x_{r_1,t_c}}{ \dot{x}_{1, t_c}}+\frac{\dot{x}_{r,t_c}}{ \dot{x}_{1, t_c}}\right)+\frac{1}{\hat{m}_{e} s}\right)\right|_{t=t_s} +\frac{1}{\omega_r^2}\left.\left(\frac{d^{2} \hat{m}_{e, s}(t)}{d t^{2}} \frac{x_{r_1, t_{c}}}{\dot{x}_{1, t_c}}+\frac{d {\hat{m}_{e, s}}(t)}{d t} \frac{\dot{x}_{r_1, t_c}}{\dot{x}_{1, t_c}}\right)\right|_{t=t_{s}}\right),\\
\it{CM}_{a} =& \frac{m}{m+m_r}\left(\left.\mathcal{L}^{-1} \left(\frac{\hat{m}_{e, s}}{s}\left(s \frac{x_{r_1,t_c}}{ \dot{x}_{1, t_c}}+\frac{\dot{x}_{r,t_c}}{ \dot{x}_{1, t_c}}\right)+\frac{1}{s}\right)\right|_{t=t_s} +\frac{1}{\omega_r^2}\left.\left(\frac{d^{2} \hat{m}_{e, s}(t)}{d t^{2}} \frac{x_{r_1, t_{c}}}{\dot{x}_{1, t_c}}+\frac{d {\hat{m}_{e, s}}(t)}{d t} \frac{\dot{x}_{r_1, t_c}}{\dot{x}_{1, t_c}}\right)\right|_{t=t_{s}}\right),\\
\it{CM}_{a} =& \frac{m}{m+m_r}\left(1+\left.\left(\hat{m}_{e, s}(t) \frac{x_{r_1,t_c}}{ \dot{x}_{1, t_c}}+\int_0^t\hat{m}_{e, s}(t).dt\frac{\dot{x}_{r,t_c}}{ \dot{x}_{1, t_c}}\right)\right|_{t=t_s} +\frac{1}{\omega_r^2}\left.\left(\frac{d^{2} \hat{m}_{e, s}(t)}{d t^{2}} \frac{x_{r_1, t_{c}}}{\dot{x}_{1, t_c}}+\frac{d {\hat{m}_{e, s}}(t)}{d t} \frac{\dot{x}_{r_1, t_c}}{\dot{x}_{1, t_c}}\right)\right|_{t=t_{s}}\right),\\
\it{CM}_{a} =& \frac{m}{m+m_r}\left(1+\frac{x_{r_1,t_c}}{ \dot{x}_{1, t_c}}\left.\left(\hat{m}_{e, s}(t) + \frac{1}{\omega_r^2}\frac{d^{2} \hat{m}_{e, s}(t)}{d t^{2}}\right)\right|_{t=t_{s}}+\frac{\dot{x}_{r,t_c}}{ \dot{x}_{1, t_c}}\left.\left(\int_0^t\hat{m}_{e, s}(t).dt+\frac{1}{\omega_r^2}\frac{d {\hat{m}_{e, s}}(t)}{d t}\right)\right|_{t=t_{s}}\right).\\
\end{aligned}
$$
Substituting for $\mathcal{L}^{-1}\hat{m}_{e,s}(s)=\hat{m}_{e,s}(t)=(m_r/m)\omega_r\sin{(\omega_r t)}$ leads to
$$
\begin{aligned}
\it{CM}_{a} =& \frac{m}{m+m_r}\left(1+\frac{m_r}{m}\frac{x_{r_1,t_c}}{ \dot{x}_{1, t_c}}\left(\omega_r\sin{(\omega_r t_s)} - \omega_r\sin{(\omega_r t_s)}\right)+\frac{m_r}{m}\frac{\dot{x}_{r,t_c}}{ \dot{x}_{1, t_c}}\left(-\cos{(\omega_rt_s)}+1+\cos{(\omega_rt_s)}\right)\right).\\
\Aboxed{\it{CM}_{a} =& \frac{m}{m+m_r}\left(1+\frac{m_r}{m}\frac{\dot{x}_{r,t_c}}{ \dot{x}_{1, t_c}}\right).}
\end{aligned}
$$
\end{section}